\documentclass[journal,10pt]{IEEEtran}
\usepackage{amsmath,amsfonts,amssymb}
\usepackage{algorithm}
\usepackage{algpseudocode}
\usepackage{array}
\usepackage{textcomp}
\usepackage{stfloats}
\usepackage{url}
\usepackage{verbatim}
\usepackage{graphicx}
\usepackage{cite}
\usepackage{multirow}
\usepackage{color}
\usepackage[table]{xcolor}
\usepackage{caption}
\usepackage{subcaption}
\usepackage{graphicx}

\usepackage{makecell}

\begin{document}
\title{LR-FHSS Transceiver for Direct-to-Satellite IoT Communications: Design, Implementation, and Verification}
\author{Sooyeob~Jung,~Seongah~Jeong,~Jinkyu~Kang,~Gyeongrae~Im,\\ \ \ \ \ \ \ ~Sangjae~Lee,~Mi-Kyung~Oh,~Joon~Gyu~Ryu,~and~Joonhyuk~Kang
\thanks{The work of Sooyeob Jung was supported by Institute of Information \& communications Technology Planning \& Evaluation (IITP) grant funded by the Korea government (MSIT) (No.2020-0-00843, Development of low power satellite multiple access core technology based on LEO cubesat for global IoT service).} 
\thanks{The work of Seongah Jeong was supported by the National Research Foundation of Korea (NRF) grant funded by the Korea government (MSIT) (No.2023R1A2C2005507).} 
\thanks{The work of Jinkyu Kang was supported by the NRF grant funded by the Korea government (MSIT) (No.2021R1F1A1050734).} 
\thanks{The work of Joonhyuk Kang was supported by the MSIT, Korea, under the ITRC (Information Technology Research Center) support program (IITP-2024-2020-0-01787) supervised by the IITP.}
\thanks{Sooyeob Jung is with the School of Electrical Engineering, Korea Advanced Institute of Science and Technology (KAIST), Daejeon 34141, and with the Satellite Communication Infra Research Section, Electronics and Telecommunications Research Institute (ETRI), Daejeon 34129, South Korea (Email: jung2816@kaist.ac.kr).} 
\thanks{Seongah Jeong is with the School of Electronics Engineering, Kyungpook National University, Daegu 41566, South Korea (Email: seongah@knu.ac.kr).}
\thanks{Jinkyu Kang is with the Department of Information and Communications Engineering, Myongji University, Gyeonggi-do 17058, South Korea (Email: jkkang@mju.ac.kr).}
\thanks{Gyeongrae Im, Sangjae Lee, Mi-Kyung Oh, and Joon Gyu Ryu are with the Satellite Communication Infra Research Section, ETRI, Daejeon 34129, South Korea (Email: \{imgrae, leestrike, ohmik, jgryurt\}@etri.re.kr)}
\thanks{Joonhyuk Kang is with the School of Electrical Engineering, KAIST, Daejeon 34141, South Korea (Email: jhkang@ee.kaist.ac.kr).}}

\markboth{Submitted in IEEE Internet of Things Journal}
{}

\maketitle

\begin{abstract}
This paper proposes a long range-frequency hopping spread spectrum (LR-FHSS) transceiver design for the Direct-to-Satellite Internet of Things (DtS-IoT) communication system. The DtS-IoT system has recently attracted attention as a promising nonterrestrial network (NTN) solution to provide high-traffic and low-latency data transfer services to IoT devices in global coverage. In particular, this study provides guidelines for the overall DtS-IoT system architecture and design details that conform to the Long Range Wide-Area Network (LoRaWAN) [\ref{Previous1}]. Furthermore, we also detail various DtS-IoT use cases. Considering the multiple low-Earth orbit (LEO) satellites, we developed the LR-FHSS transceiver to improve system efficiency, which is the first attempt in real satellite communication systems using LR-FHSS. Moreover, as an extension of our previous work [\ref{COMML}] with perfect synchronization, we applied a robust synchronization scheme against the Doppler effect and co-channel interference (CCI) caused by LEO satellite channel environments, including signal detection for the simultaneous reception of numerous frequency hopping signals and an enhanced soft-output-Viterbi-algorithm (SOVA) for the header and payload receptions. Lastly, we present proof-of-concept implementation and testbeds using an application-specific integrated circuit (ASIC) chipset and a field-programmable gate array (FPGA) that verify the performance of the proposed LR-FHSS transceiver design of DtS-IoT communication systems. The laboratory test results reveal that the proposed LR-FHSS-based framework with the robust synchronization technique can provide wide coverage, seamless connectivity, and high throughput communication links for the realization of future sixth-generation (6G) networks.
\end{abstract}


\IEEEpeerreviewmaketitle

\section{Introduction}
Due to the rapid growth of Internet of Things (IoT), especially with the outbreak of the COVID-19 pandemic, the evolution of fifth-generation (5G) wireless communication networks has accelerated towards future sixth-generation (6G) wireless networks [\ref{6G1}]-[\ref{6G4}]. Future 6G networks need to meet the new requirements for global coverage and connectivity, as well as high data rates, high reliability, low latency, and energy efficiency, which have attracted attention from existing terrestrial communications. To this end, the development and integration of nonterrestrial networks (NTNs) is indispensable, which has encouraged the revolution of Direct-to-Satellite IoT (DtS-IoT) systems, so-called Internet of Remote Things [\ref{DTS1}]-[\ref{DTS3}]. Among the various types of satellites, such as geostationary-Earth orbit (GEO), medium-Earth orbit (MEO), highly-elliptical orbit (HEO) and low-Earth orbit (LEO), LEO satellites are considered the most suitable for future IoT applications due to their low latency, cost-effectiveness, and ease-of-deployment [\ref{LEO1}].

DtS-IoT offers low dependence on ground base stations as well as simplifies communications to remote areas, especially where deploying terrestrial infrastructure is challenging. To realize a scalable and energy-efficient DtS-IoT system, orthogonal frequency-division multiplexing (OFDM)-based waveforms specified in the third generation partnership project (3GPP) [\ref{NBIoT}] and low-power wide area network (LPWAN) technologies [\ref{LPWAN}]-[\ref{Ingenu}] have been studied. This paper focuses on LPWAN technologies available in the industrial, scientific, and medical (ISM) band that can be used for any purpose without a license in most countries.

To overcome the capacity limitation of existing LPWAN technologies, which mostly focus on terrestrial IoT communications under LEO satellite channel environments [\ref{ETRIJ}]-[\ref{LEO}], Semtech [\ref{Previous6}] has proposed a long range-frequency hopping spread spectrum (LR-FHSS) transmission technology. To the best of our knowledge, there have been only a limited number of studies conducted on LR-FHSS-based LPWAN systems. The existing research [\ref{Previous2}]-[\ref{Previous8}] has shown that the LR-FHSS provides a substantial improvement in capacity compared to other LPWAN transmission schemes, particularly for Long Range (LoRa) communications systems [\ref{LoRa}]. This demonstrates that the LR-FHSS is well-suited for global coverage of DtS-IoT transmission. Most of the previous works [\ref{Previous2}]-[\ref{FHSS_add}] explore performance analysis in terms of the throughput and the outage probability. This is mainly due to the limited information about the LR-FHSS physical-layer specifications in the LoRaWAN [\ref{Previous1}]. 

\begin{figure*}[t]
\centering
    \includegraphics[width=0.9\textwidth]{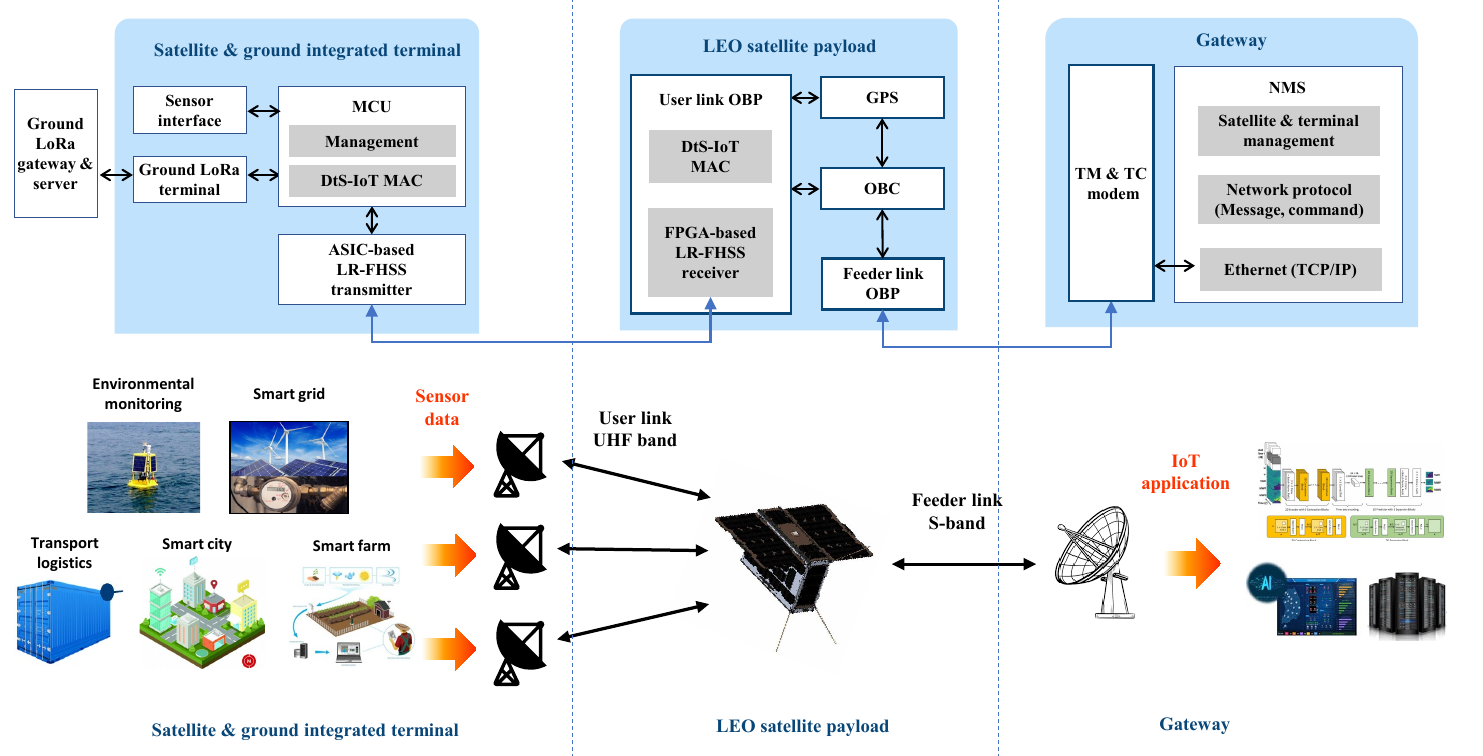}
    \caption{Overall DtS-IoT system architecture using the LR-FHSS transmission.}
\end{figure*}

Our previous work [\ref{COMML}] suggests an LR-FHSS transceiver structure that is compliant with the LoRaWAN specifications [\ref{Previous1}] published by LoRa Alliance and Semtech’s application notes [\ref{Previous6}] as the first trial. The proposed transceiver structure is designed with the perfect synchronization assumption, whose performances are verified in terms of the miss-detection probability and packet error probability (PER). However, actual LEO satellite channel environments involve Doppler effects, symbol timing offset (STO), carrier frequency offset (CFO), phase offset, and co-channel interference (CCI), making synchronization critical for the desired performance of DtS-IoT applications. The design of LR-FHSS-compatible synchronization procedures remains a problem, calling for effective solutions that consider actual LEO satellite impairments. The design complexity of LR-FHSS-compatible synchronization is exacerbated by the fact that the LR-FHSS foresees frequency-hopping channels up to 3120, demanding a specific frequency-hopping pattern for collision avoidance. This adds stress on the optimized design of signal detection, header reception, and payload reception, where the signal detector requires a structure that can simultaneously receive up to 3120 frequency-hopping channels blindly in terms of the frequency and time domains. To approach perfect synchronization, the synchronization algorithm must be designed to overcome the following channel impairment conditions: an initial STO up to 1/4, sampling frequency offset (SFO) up to 80 ppm, CFO up to 5/6 of the symbol rate, Doppler rate up to 400 Hz/sec, and CCI up to 40 percent.

For practicality and reliability, we developed a feasible solution of design and implementation for a robust DtS-IoT communication framework based on an LR-FHSS transceiver against the LEO channel impairments that conforms to the LoRaWAN standards [\ref{Previous1}]. Specifically, we developed the Gaussian minimum shift keying (GMSK)-based LR-FHSS transceiver including a robust synchronization algorithm with a GMSK symbol mapping scheme suitable for LR-FHSS transmission [\ref{COMML}]. Then, based on the proposed LR-FHSS transceiver design, we built a DtS-IoT framework consisting of satellite and ground integrated terminals, LEO satellite payload, and gateway, as illustrated in Fig. 1. The specific contributions of this paper are summarized as follows:

\begin{itemize}
\item \textit{Overall DtS-IoT system architecture design}: The system architecture and design details for the DtS-IoT systems with the LR-FHSS transceiver are suggested with various use cases and applications. Considering the multiple satellites in orbit, a standard-compliant LR-FHSS transceiver with LoRaWAN specification [\ref{Previous1}] was designed to support the high throughput for spectral efficiency. In addition, we explore and analyze the channel impairments that should be considered for LEO satellite communications.

\item \textit{Robust synchronization algorithm development against the LEO channel impairments}: A robust synchronization algorithm to withstand the LEO channel impairments was developed for LR-FHSS transmission with thousands of frequency hopping channels. Signal detection is included for the simultaneous reception of numerous frequency hopping signals and enhanced soft-output-Viterbi-algorithm (SOVA) for the header and payload receptions. In detail, the specific channel impairments such as an initial STO up to 1/4, SFO up to 80 ppm, CFO up to 5/6 of the symbol rate, Doppler rate up to 400 Hz/sec, and CCI up to 40 percent can be considered.

\item \textit{Implementation}: The implementation details of testbeds using an application-specific integrated circuit (ASIC) chipset and a field-programmable gate array (FPGA) are illustrated for verification. Laboratory test results reveal that the proposed LR-FHSS transceiver design can provide robustness against the Doppler effect and CCI caused by LEO satellite channel environments, enabling wide coverage, seamless connectivity and high throughput communication links for future 6G networks. In particular, DtS-IoT systems with unslotted and slotted Aloha multiple access schemes support throughputs of over 0.18 and 0.37, respectively. To the best of our knowledge, this is the first proposal for realizing an LEO satellite communication system using LR-FHSS. Moreover, by presenting the expected DtS-IoT use cases with the corresponding key performance indicators (KPIs), we can guide the readers in developing the DtS-IoT system.
\end{itemize}

The remainder of this paper is organized as follows. Section \ref{sec:Section 2} introduces related works and suggests the proposed DtS-IoT system architecture with potential use cases. The robust synchronization algorithm for the LR-FHSS transmission is detailed in Section \ref{sec: Section3}. In Section \ref{sec: Section4}, implementation and testbeds to integrate the system components are described, whose performances are verified via laboratory tests. Finally, conclusions and future works are summarized in Section \ref{sec:multiple_orbits}.

\section{DtS-IoT System Architecture Design}\label{sec:Section 2}
In this section, we establish the overall system architecture for DtS-IoT communication based on an LR-FHSS transceiver and the potential DtS-IoT use cases with the associated KPIs. In addition, to achieve the quality of service (QoS) of DtS-IoT communications, we explore the challenging issues that must be addressed.

\subsection{Previous Works}
Due to the limited availability of information on the LR-FHSS physical-layer in the LoRaWAN specifications [\ref{Previous1}], only a few studies have been conducted: 
[\ref{COMML}], [\ref{Previous2}]-[\ref{Previous8}]. Most of them [\ref{Previous2}]-[\ref{Previous5}] mainly focus on throughput and outage probabilities. In particular, [\ref{Previous2}] compares the network capacity performance of LR-FHSS and LoRa under no fading channel conditions. In [\ref{Previous3}], the success probability for packet delivery is mathematically investigated  under noise-free channel conditions, including path-loss and Rician fading in LR-FHSS systems. Maleki \textit{et al.} [\ref{Previous4}] derive a closed-form expression for outage probability under realistic channel conditions, considering path-loss and Nakagami-m fading. Furthermore, [\ref{Previous5}] explores a shadowed-Rician fading model, resembling the actual satellite channel, to analyze the outage probability of device-to-device-aided LR-FHSS schemes. In [\ref{Previous2}]-[\ref{Previous5}], the suitability of LR-FHSS for DtS-IoT transmission is demonstrated, showcasing significant capacity improvements compared to LPWAN transmission schemes. Recently, several signal detection techniques have been proposed based on the LR-FHSS physical layer: [\ref{FHSS_add}]-[\ref{Previous8}]. Fraire \textit{et al.} [\ref{FHSS_add}] developed the heuristic headerless signal detection method based on the integer linear program for extreme conditions. In [\ref{Previous7}], the interference blanking technique is suggested to increase the number of LR-FHSS signals received simultaneously. In addition, [\ref{Previous8}] proposes a new method to dynamically adjust the frequency hopping sequence and improve the reception rate of LR-FHSS by developing a reception algorithm. However, [\ref{FHSS_add}]-[\ref{Previous8}] only deal with signal detectors, which are part of the overall receiver, and perform algorithm verification in an imperfect LEO satellite channel environment.

In our previous work [\ref{COMML}], a novel transceiver design is proposed for LR-FHSS-based DtS-IoT systems. The header reception is assumed to be possible via header repetition, as in the standard [\ref{Previous1}], since LR-FHSS networks can manage the collision rate among multiple end devices (EDs). Specifically, [\ref{COMML}] proposes a transmitter structure based on the standardized physical-layer specifications of LoRaWAN [\ref{Previous1}] by modifying the preamble insertion introduced incorrectly in the existing LR-FHSS studies [\ref{Previous2}]-[\ref{Previous5}] and developing the new time-on-air (TOA) calculation. Moreover, for the first time, besides commercial products, we designed the receiver structure with blind LR-FHSS that can simultaneously receive up to 3120 frequency hopping channels, which can be mixed in terms of the frequency and time domains.

\subsection{Overall System Architecture and Potential Use Cases}
For end-to-end transmission, the DtS-IoT system, in general, consists of the terminal that collects data from IoT sensors and transmits it to the satellite via uplink, the LEO satellite payload that stores data transmitted from the terminal and transmits it to the gateway via downlink, and the gateway that receives data from the LEO satellite payload and processes it. The goal of the DtS-IoT system is to support seamless data delivery services with global coverage and connectivity. Currently, only high-latency services can be supported through multiple LEO satellites operating in orbit. In the near future, satellite operators such as SpaceX, Amazon, and Telesat [\ref{LEO1}] plan to support low-latency and seamless services by launching thousands of LEO satellites, installing multiple ground base stations or establishing inter-satellite links. The ultimate aim of DtS-IoT communications is to develop a waveform dedicated to LEO satellite communication that can support high throughput. LR-FHSS transmission technology has recently been proposed by Semtech [\ref{Previous6}] that has a high network capacity and collision robustness in long-range and large-scale communication scenarios. In this paper, we construct a comprehensive DtS-IoT system using LR-FHSS transmission and reception, and then explain the roles and interfaces of essential components in detail.

Fig. 1 shows the overall DtS-IoT system architecture, which is composed of the satellite and ground integrated terminal, LEO satellite payload and gateway, with the LR-FHSS transmission and reception. First, the satellite and ground integrated terminals receive the data collected from the multiple IoT sensors for various use cases, such as environmental monitoring, smart grid, transport logistics, smart city, and smart farm [\ref{Usecase1}]-[\ref{Usecase4}]. Through the sensor interface and monitoring and control unit (MCU) modules in the terminal, the collected data is transmitted to the LEO satellite payload from the ASIC-based LR-FHSS transmitter in the ultra high frequency (UHF) band, i.e., 940 MHz. This frequency band can be selected by considering the output power and channel occupancy time in the unlicensed band. It also includes the module for the LoRa transmission and reception to link the ground gateway and server. In the LEO satellite payload, which is composed of the user link and feeder link onboard processors (OBPs), onboard computer (OBC), and global positioning system (GPS), the data transmitted from the satellite and ground integrated terminals is demodulated by the FPGA-based LR-FHSS receiver in the user link OBP. The demodulated data is stored in the OBC, while the LEO satellite payload flies in the predetermined orbit until it can communicate with the gateway. When the connection with the gateway is possible, the stored data is transferred from the feeder link OBP to the gateway in the S-band for high-speed data transfer. For data transmission on the feeder link, the telemetry (TM) and telecommand (TC) modem, i.e., CORTEX CRT modem, can be applied between the LEO satellite payload and the gateway. In the gateway, the network management system (NMS) receives the demodulated data from the TM and TC modem via ethernet and processes it in the network protocol module. The network protocol module can receive the collected data and send and receive TM and TC messages, while sending beacon setting information. Also, the satellite and terminal management module can manage and store satellite and terminal information of the DtS-IoT. Ultimately, the IoT data stored in the gateway can be used depending on the application, e.g., to predict marine climate through machine learning-based big data analysis.

According to the utilization of collected IoT data, target use cases with their respective KPIs are summarized in Table I [\ref{Usecase1}]-[\ref{Usecase4}]. The requirements of the use cases are defined in the same way as the European telecommunications standards institute (ETSI) technical report (TR) 103 435 [\ref{Usecase1}] in terms of data rate, power, and latency. For data rate, environmental monitoring, smart grid, transport logistics and smart farm that require simple IoT data collection can support low data rates of 1 kbps or less. However, smart cities require complex processing of data collected from various IoT terminals, requiring high data rates of up to 10 kbps. In terms of power consumption, IoT terminals equipped with portable batteries require low power in all use cases, as power is supplied at approximately 30 volts or less. Latency is not an important factor for environmental monitoring and smart farms which require long-term data collection. For example, environmental monitoring can be used to measure air quality, soil quality, ocean conditions, and animal tracking, which requires high hardware environmental resistance to factors such as temperature and humidity, fully functional GPS, and energy harvesting capabilities. Smart farms use IoT terminals for water pump control, sensors for oxygen, temperature and humidity, hydraulic valves, and level meters; therefore, their latencies must be quite relaxed, or between a few seconds and a few minutes. Conversely, the remaining use cases utilize IoT data in urgent situations of short-or mid-term, thus requiring a latency of 100 ms to 1000 ms. The smart grid uses IoT as a fault passage indicator, low voltage sensors, transformer status monitoring and alarm detection. Transport logistics uses the technology to support contents such as shipment tracking sensors, microprocessors, wireless connectivity, and minuscule identification devices. For smart cities, it supports the content such as the traffic counting detectors, inductive loop detectors for vehicle identification, highly accurate GPS and sensors for CO2, fog, and ultrasonic monitoring.

\begin{table}[t]
    \caption{Target DtS-IoT use cases with their respective KPIs}\label{Table1}
    \centering
    \begin{tabular}{| p{2cm} | p{1.5cm} | p{1.2cm} | p{2cm} |}
     \hline
     \centering \bf{Use Cases} & \centering \bf{Data rate} & \centering \bf{Power} & \ \ \ \ \ \ \bf{Latency} \\
     \hline
     \centering Environmental monitoring & \centering \multirow{2}{*}{$\le 1$ kbps} & \centering \multirow{2}{*}{Low} & \multirow{2}{*}{Quite relaxed} \\   
     \hline
     \centering Smart \\ grid & \centering \multirow{2}{*}{$\le 1$ kbps} & \centering \multirow{2}{*}{Low}  & \multirow{2}{*}{$100 \sim 1000$ ms} \\
     \hline     
     \centering Transport logistics & \centering \multirow{2}{*}{$\le 1$ kbps} & \centering \multirow{2}{*}{Low} & \multirow{2}{*}{$100 \sim 1000$ ms}\\
     \hline       
     \centering Smart \\ city & \centering \multirow{2}{*}{$\le 10$ kbps} & \centering \multirow{2}{*}{Low} & \multirow{2}{*}{$100 \sim 1000$ ms} \\
     \hline   
     \centering Smart \\ farm & \centering \multirow{2}{*}{$\le 1$ kbps}& \centering \multirow{2}{*}{Low} & Few seconds  $\sim$  minutes\\
     \hline       
    \end{tabular}
\end{table}

\begin{figure}[t]\label{Tx}
\centering
    \includegraphics[width=\columnwidth]{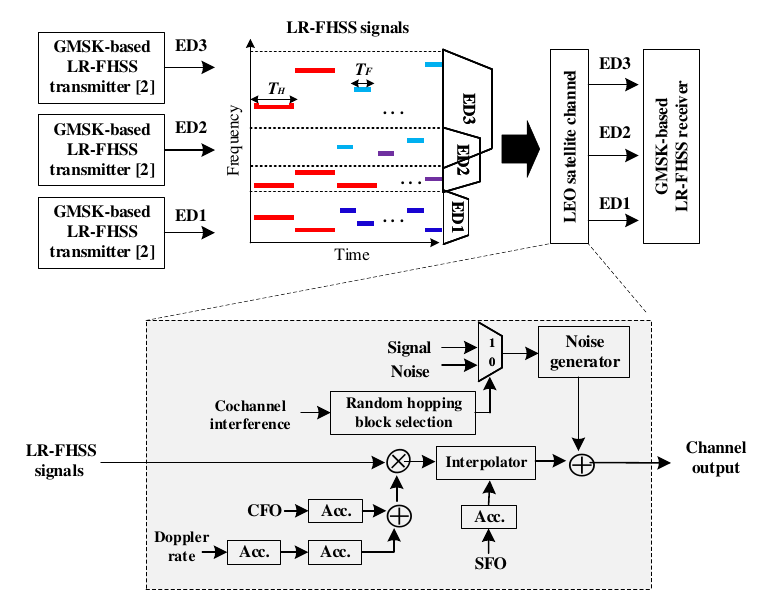}
    \caption{Block diagram of LR-FHSS transmission with the LEO satellite channel model.}
\end{figure}

\subsection{LR-FHSS Transceiver Design}
Fig. 2 shows a block diagram of the LR-FHSS transmission with the LEO satellite channel model. Since the details of the LR-FHSS transmitter design, including the header and payload transmission through frequency hopping, is introduced in our previous work [\ref{COMML}], here, we briefly summarize the major components. By applying GMSK symbol mapping suitable for LR-FHSS, the hopping blocks of 114 symbols for header fragments and 50 symbols for payload fragments are generated. In the LR-FHSS transmitter output, a total of $N_H + N_F$ hopping blocks, considering the number of header replicas ${N_H}$ with the header duration ${T_H}$ and the number of payload fragments ${N_F}$ with the payload duration ${T_F}$, are transmitted by switching frequency channels based on the hopping channel and pattern. In Fig. 2, the header and payload fragments, the outputs of three EDs, are simultaneously transferred in the frequency-time domain using frequency hopping. In LR-FHSS networks, the collision rate between the multiple EDs can be reduced by managing the spectrum used by each ED. To efficiently operate DtS-IoT communication systems, the LR-FHSS network needs to allocate a clean spectrum for EDs with a high QoS or with low transmit power. For example, supposing that ED1 transfers the LR-FHSS signal to require less transmit power and a higher QoS than ED2 and ED3, the overlapped frequency band for ED2 and ED3 as well as the clean frequency band for ED1 can be allocated.

The LR-FHSS receiver structure consisting of signal detection and decoding blocks is proposed in [\ref{COMML}] with perfect synchronization. The design of the LR-FHSS compatible synchronization procedure remains an open problem, which is inevitable in real applications, but challenging for implementation. In this paper, we suggest a novel transceiver structure including the robust synchronization algorithm to compensate for actual LEO satellite channel effects, such as the Doppler effect, STO, CFO, phase offset, and CCI, which is explained in detail in Sec. III.

\subsection{Challenging Issues}
In this section, we provide an overview of the challenging issues encountered during the implementation of the LR-FHSS-based DtS-IoT system, and possible solutions to these challenges are proposed in Table II.

\begin{table}[t]
    \caption{Impairments causes and their solutions}\label{Table4}
    \centering
    \begin{tabular}{| p{1.5cm} | p{6.4cm} | }
     \hline
     \centering \bf{Impairments} & \ \ \ \ \ \ \ \ \ \ \ \ \ \ \ \ \ \ \ \ \ \ \ \ \  \bf{Solutions} \\
     \hline
     \centering Doppler effect & comparison of the several Doppler candidates (11 candidates in this paper), Doppler tracking in header SOVA\\  
     \hline     
     \centering STO & Symbol timing estimation using a syncword  \\
     \hline     
     \centering \multirow{2}{*}{CFO} & Coarse CFO estimation in signal detection, fine CFO estimation, CFO tracking in header SOVA  \\   
     \hline   
     \centering Phase \ \ offset & \multirow{2}{*}{Phase estimation, phase tracking in header SOVA}  \\   
     \hline        
     \centering \multirow{2}{*}{CCI} & Interleaving, \ convolutional encoding, \ frequency hopping \\
     \hline
     \centering Signal detection & \multirow{2}{*}{Signal detection in frequency domain}  \\   
     \hline
     \centering Time difference & \multirow{2}{*}{Signal detection in frequency domain} \\
     \hline
    \end{tabular}
\end{table}

\subsubsection{Doppler Effect}
In LEO satellite communications, significant and time-variant frequency shifts can occur due to the Doppler effect, depending on factors such as carrier frequency, satellite altitude, orbit, and coverage assigned to each LEO satellite [\ref{ETRIJ}]-[\ref{LEO}]. The Doppler effect has a notable impact on coherent demodulation, necessitating compensation for reliable communications. As shown in Fig. 2, the Doppler effect can be emulated by considering that the Doppler rate passes through two accumulators. The output of the first accumulator represents a Doppler shift, with the Doppler rate typically affecting the narrowband around 488 Hz. This paper considers Doppler rates up to 400 Hz/sec. Initially, the Doppler effect can be estimated through performance comparison among several Doppler candidates; for instance, our framework employs 11 candidates. Subsequently, Doppler tracking is performed in the header SOVA block.

\subsubsection{STO}
The STO arises from the long-term instabilities of the receiver oscillator, which generates the receiver sampling clock [\ref{DVB}]. STO can lead to inter-symbol interference (ISI), significantly degrading reception performance. To address STO in the channel model, SFO is accumulated for each sample and converted to the sampling phase, which is then applied to the LR-FHSS signal via an interpolator, as illustrated in Fig. 2. The precision of the oscillator, typically limited by its cost, is usually capped at a maximum of 10 ppm; for instance, in our architecture, up to 80 ppm is considered. For symbol timing estimation, the initial STO and SFO are estimated through cross-correlation using a syncword.

\subsubsection{CFO and Phase Offset}
The CFO can be attributed to various factors leading to performance degradation, including receiver oscillator instabilities and Doppler effects [\ref{DVB}]. CFO alters the frequency of the received signal, potentially causing mismatches and resulting in inter-carrier interference (ICI). By accumulating CFO for each sample, the calculated phase component is applied to the LR-FHSS input considering the Doppler effect. In our proposed framework, we consider CFO compensation up to 5/6 of the symbol rate which can be achieved through several stages, including coarse estimation, fine estimation, and tracking. Similarly to CFO, phase offset can also be compensated for through phase estimation and tracking.

\subsubsection{CCI}
For DtS-IoT systems, the utilization of unslotted Aloha-based multiple access [\ref{Unslotted}] is advantageous for increasing network capacity but can also lead to CCI among multiple EDs. In the channel model, CCI can be introduced alongside noise by randomly selecting the overlapping ratio of hopping blocks from different EDs within the total frame length. Depending on the coding rate, the impact of CCI may vary. In our framework, it is observed that up to 40 percent of hopping block overlaps can occur without encountering an error floor. During frequency hopping transmission, CCI can be mitigated through spectrum management of the LR-FHSS network. However, occasional collisions between LR-FHSS packets can be addressed through interleaving and convolutional encoding. Furthermore, the signal detection block is capable of detecting frequency hopping signals from multiple EDs at different output times [\ref{COMML}].

\begin{figure*}[t]
     \centering
        \includegraphics[width=0.9\textwidth]{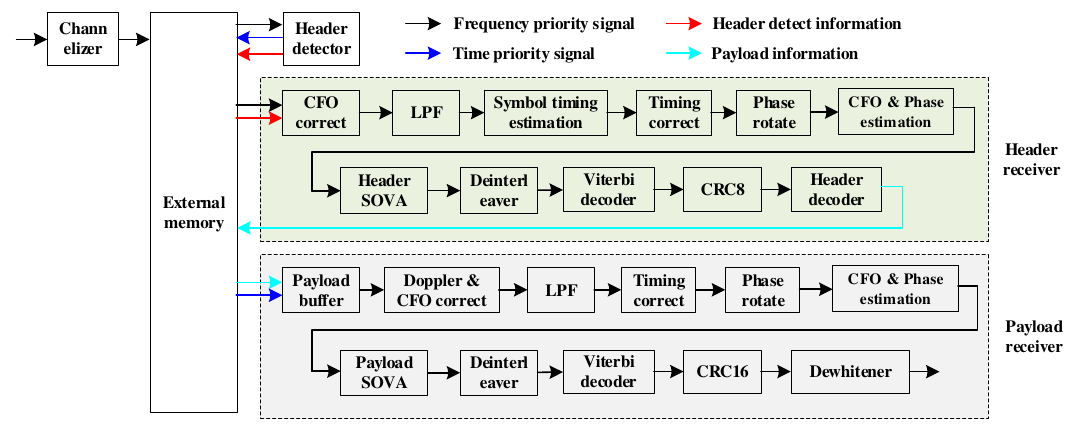}
        \caption{Structure of the GMSK-based LR-FHSS receiver for synchronization and header decoding.}
        \label{sync}        
\end{figure*}

\section{Robust Synchronization for DtS-IoT Communications}\label{sec: Section3}
In this section, we propose the structure of the GMSK-based LR-FHSS receiver, as shown in Fig. 3, which is composed of the signal detector, header receiver, and payload receiver. For receiving the LR-FHSS signal, a robust synchronization algorithm that can withstand real LEO satellite channel environments is essential, whose details and performances are described in the following subsections.

\subsection{Signal Detector}
As shown in Fig. 3, the LR-FHSS signal detector structure is composed of a channelizer, external memory, and header detector. In the signal detector, the channelizer can simultaneously receive up to $N_{CF} = 3120$ LR-FHSS signals available in the operating channel width (OCW) of 1523 kHz by using the FFT operation with windowing. The FFT output of the channelizer is transferred to the external memory per frequency bin. The external memory is divided into the memory regions, where the data is stored in both the frequency and time axis. The FFT output corresponds to all frequency signals at one sampling point, which is referred to as the frequency priority signal. First, the frequency priority signal is stored in the corresponding memory region. In the time domain, two oversampled signals for each frequency are stored in the time order, which is called a time priority signal. After header detection, the time priority signal is stored in the corresponding memory region. To detect the header position, the frequency priority signal for each frequency is transferred to the header detector. Using the frequency priority signal, the header detector then makes a detection decision through cross-correlation. In the header detector, we can estimate the header position corresponding to the sampling point and the channel information corresponding to the frequency hopping channel. In addition, the coarse CFO can be estimated with a resolution of $976/${FFT\_size} Hz. The header position information obtained from the header detector is then transferred to external memory.

\subsection{Header Receiver} 
After detecting the LR-FHSS signal, the header receiver performs synchronization and header decoding using the frequency priority signal with the header detect information. As shown in Fig. 3, the header receiver is composed of blocks for synchronization, such as the symbol timing estimation, phase rotate, and the CFO and phase estimation, and blocks for header decoding, such as the header SOVA, deinterleaver, Viterbi decoder, CRC8 and header decoder.

\subsubsection{Symbol Timing Estimation}
For the symbol timing estimation, a header signal sequentially passed through the CFO correct and low-pass filter (LPF) blocks is the input. In the CFO correct block, the coarse CFO estimated by the signal detector is compensated in the header, and the noise outside the signal band is removed in the LPF. The filtered input signal enables precise estimation of STO, which causes ISI.

In the symbol timing estimation block, we determine the STO using the syncword information embedded in the 114-symbol header. The STO is generated by a change in sampling phase due to accumulated SFO. Specifically, the cross-correlation of sample timings is calculated before and after one symbol and an accurate sampling phase is obtained by comparing the magnitude squared values of the cross-correlation. This block utilizes two oversampled signal inputs and syncword-based correlator coefficients. The estimated sampling phase is then corrected in an interpolator.

\begin{figure}
\centering
\begin{subfigure}{0.24\textwidth}
    \includegraphics[width=\textwidth]{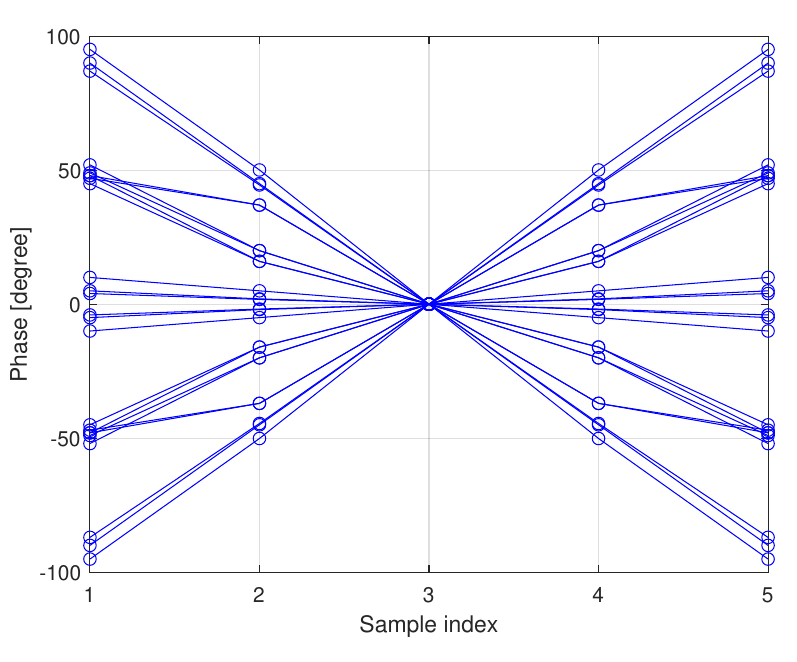}
    \caption{}
\end{subfigure}
\begin{subfigure}{0.24\textwidth}
    \includegraphics[width=\textwidth]{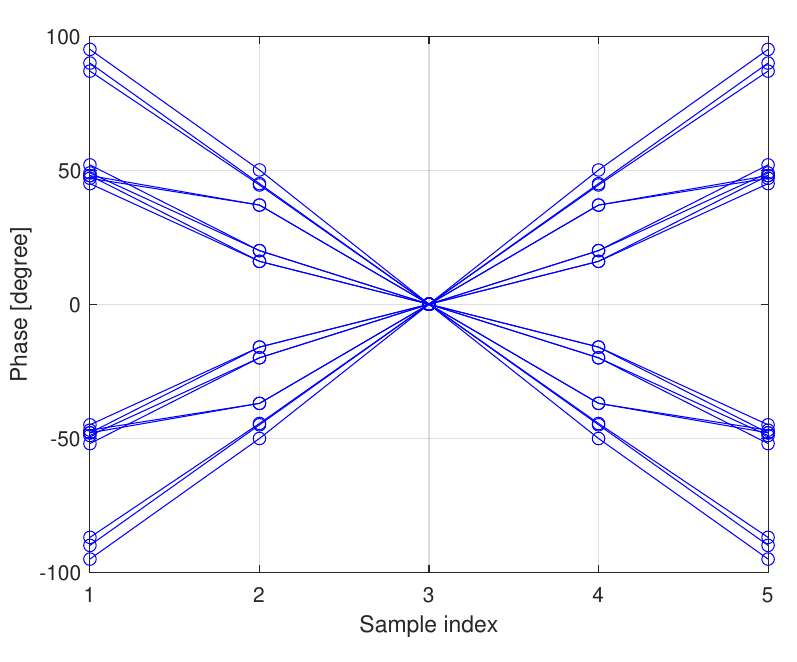}
    \caption{}
\end{subfigure}
\caption{Phase change of the syncword with (a) the full length of 32 symbols and (b) partial length of 20 symbols.}
\end{figure}

Since the GMSK signal having a constant amplitude changes only the phase component, the STO can be estimated accurately when the phase difference between the symbol intervals is sufficiently large [\ref{STR}]. Fig. 4(a) shows the phase change of the syncword with the full length of 32 symbols, and there is little phase change in some regions around the phase of ${{0^\circ}}$. When all syncword symbols are used, the performance of the symbol timing estimation is degraded. Therefore, in our framework, the cross-correlation using 20 symbols in the \{5, 6, 7, 8, 9, 10, 11, 12, 13, 14, 15, 18, 19, 20, 21, 22, 23, 24, 25, 26\}-th positions among the 32 total symbols is applied, where these 20 symbols show a large phase difference, as shown in Fig. 4(b). In this study, the maximum frequency tolerance of 31 kHz implies a frequency error of about 33 ppm for a 940 MHz carrier frequency. Assuming the maximum symbol frequency error of 66 ppm between the LR-FHSS transmitter and receiver for a total of 2714 symbols, derived from 52 payload blocks and 1 header block in the case of the longest packet, the last symbol in the packet results in a timing error of approximately 0.18 symbol length. Fig. 5 shows that the standard deviation of the proposed symbol timing estimation corresponds to a sufficient performance within 0.1 symbol less than 0.18 symbol at the SNR of 5 dB; this SNR supports the PER of $10^{-2}$ in the LR-FHSS packet transmission under AWGN [\ref{COMML}]. In addition, it shows a performance gap of about 1 dB depending on the Doppler rate from 0 to 400 Hz/sec. This performance error indicates that symbol timing estimation using syncword is accurate; therefore, additional symbol timing tracking blocks can be omitted.

\begin{figure}[t]
     \centering
        \includegraphics[width=0.8\columnwidth]{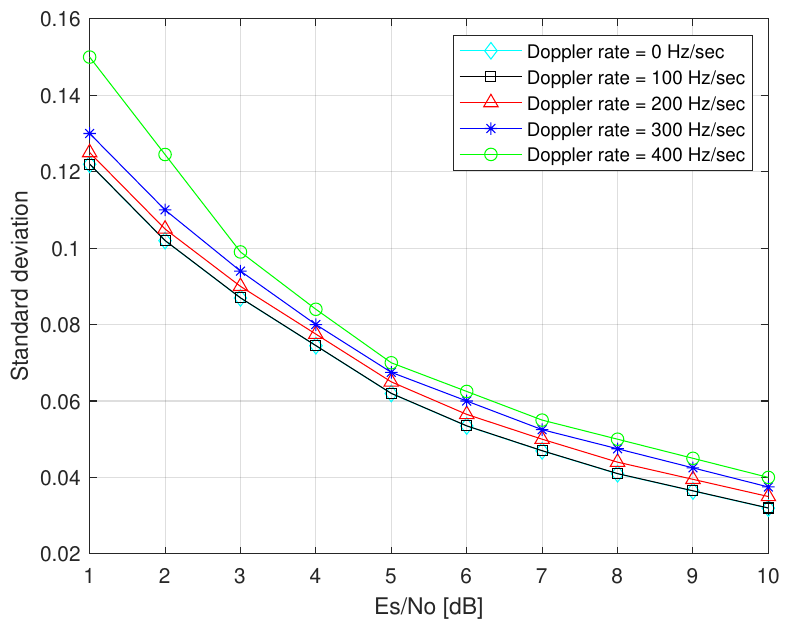}
        \caption{Standard deviations according to the Doppler rate of the proposed symbol timing estimation in the initial STO = 0.125 case.}   
\end{figure}

\subsubsection{Phase Rotate}
\begin{figure}[t]
     \centering
        \includegraphics[width=0.8\columnwidth]{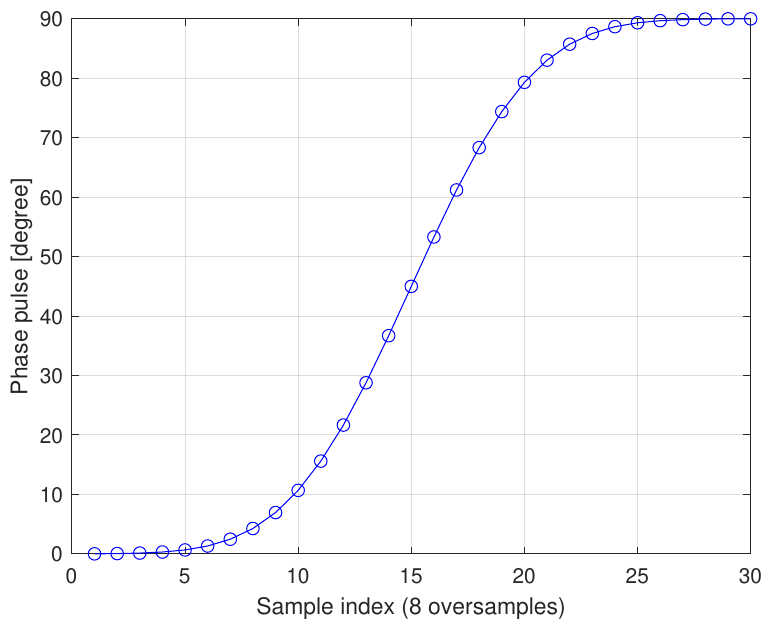}
        \caption{Phase pulse of the Gaussian filter for GMSK modulation.}    
\end{figure}

The phase of the GMSK signal has a contribution of ${ \pm {90^\circ}}$ from each symbol as a function of the previous transmitted symbol. For soft bit calculation through Trellis search [\ref{GMSK1}], [\ref{GMSK2}], eight Trellis states are required when transmitting binary symbols. If the phase of each symbol is accumulated and adjusted by ${ \pm {90^\circ}}$, the phase state by the previous symbol becomes $\{ 0^\circ ,180^\circ \}$ for an even number of symbols and $\{  - 90^\circ ,90^\circ \}$ for an odd number of symbols. Finally, four Trellis states can be obtained through phase rotation. The process for the phase rotate can be expressed as,
\begin{align}
\phi \left( {t,{\bf{a}}} \right) = \sum\nolimits_i {{a_i}q\left( {t - i{T_s}} \right)},\label{phase_pulse}
\end{align}
where $q(t)$ is the phase pulse, $a_i$ is the information bit of ${ \pm 1}$, and $T_s$ is the period of the input symbol. As shown in Fig. 6, the phase pulse becomes ${90^\circ}$ after a sufficient time. If the equation (\ref{phase_pulse}) is divided into the input symbols having a response of ${\pi \mathord /  2}$ and having a response of a phase shift interval, (\ref{phase_pulse}) can be reformulated as,
\begin{align}
\phi \left( {t,{\bf{a}}} \right) = \sum\limits_{i = k - 1}^k {{a_i}q\left( {t - i{T_s}} \right)}  + \frac{\pi }{2}\sum\limits_{i = 0}^{k - 2} {{a_i}}.\label{phase_pulse2}
\end{align}
Here, the two symbols are used as the length of the phase transition section. The first term of the equation (\ref{phase_pulse2}) is determined from the latest two symbols of $\left\{ {{a_{k - 1}},{a_k}} \right\}$, and the second term is obtained by the accumulated values from the first symbol to the symbol of ${a_{k - 2}}$.

From (\ref{phase_pulse2}), the number of Trellis states for the Viterbi detector is two for the first term and four for the second term, the total of which is eight because each of the four phases has two states. To reduce the number of Trellis states for low-complexity [\ref{GMSK2}], by adding $\frac{\pi }{2}\left( {k - 2} \right)$ to (\ref{phase_pulse2}), we have
\begin{align}
\varphi \left( {t,{\bf{a}}} \right) = & \ \phi \left( {t,{\bf{a}}} \right) + \frac{\pi }{2}\left( {k - 2} \right) \nonumber \\
= &\sum\limits_{i = k - 1}^k {{a_i}q\left( {t - i{T_s}} \right)}  + \pi \sum\limits_{i = 0}^{k - 2} {\widetilde {{a_i}}}, \label{phase_pulse3}
\end{align}
where $\widetilde {a_i}$ is defined as $1 \ {\rm{for}}\; {a_i} = 1$ and $0 \ {\rm{for}}\; {a_i} = -1$. By using (\ref{phase_pulse3}), we can reduce the number of Trellis states to four. When the channel noise is independent between symbols, the operation of changing the phase of the received signal by $\frac{\pi }{2}\left( {k - 2} \right)$ does not affect the characteristics of noise; therefore, there is no change in reception performance due to this operation. The forward and backward Trellis searches using (\ref{phase_pulse3}) are shown in Fig. 7, where ${b_k}$ is defined as
\begin{align}
{b_k} = \sum\limits_{i = 0}^{k - 2} {\widetilde {{a_i}}} \,\,\, {\rm{and}} \,\,\, {\widetilde {{b_k}}} = {b_k}\bmod \,2.
\end{align}
When denoting the received baseband signal as $r_k$, the path metric in each state $s$ of the Trellis search is calculated as
\begin{align}
\lambda _k^{(s)} = \mathop {\max }\limits_{t \in \bf{S}} \left( {\lambda _{k - 1}^{(t)} + {\mathop{\rm Re}\nolimits} \left\{ {{r_k} \cdot {e^{j\left( {k\frac{\pi }{2} - {\varphi _p\left( {k-1} \right)} - \varphi \left( {k{T_s},{\bf{a}}(s)} \right)} \right)}}} \right\}} \right),\label{phase_pulse4}
\end{align}
where $\bf{S}$ is the set of the previous states transitioning to state $s$, ${\varphi _p\left( {k} \right)}$ is the phase of the channel estimation, and ${{\bf{a}}(s)}$ is a sequence of symbols transitioning to state $s$. 

\begin{figure}
\centering
\begin{subfigure}{0.23\textwidth}
    \includegraphics[width=\textwidth]{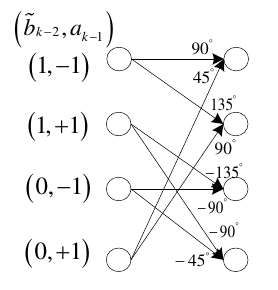}
\end{subfigure}
\begin{subfigure}{0.23\textwidth}
    \includegraphics[width=\textwidth]{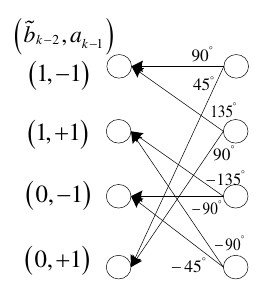}
\end{subfigure}
\caption{(a) Forward and (b) backward Trellis diagram of SOVA.}
\end{figure}

\subsubsection{CFO \& Phase Estimation}
This block estimates the CFO and phase using the syncword information of the header. First, the phase estimation ${\varphi _p\left( {k} \right)}$ is calculated by the cross-correlation between the filter coefficient $c_k$ and the input signal $r_k$, which is given as ${\varphi _p\left( {k} \right)} = {\tan ^{ - 1}}\left( {\sum\nolimits_{k = 0}^{N - 1} {{c_k}{r_k}} } \right)$, where ${{\bf{c}}_k} = \left[ {{c_0},{c_1}, \cdots ,{c_{N - 1}}} \right]$ are the filter coefficients based on the synchronous word of the hexadecimal 0x2C0F7995. 

For the CFO estimation, the result of $M$-point FFT operation with windowing of the length $N$ can be written as ${{{X}}_m} = \sum\nolimits_{k = 0}^{N - 1} {{c_k}{r_k}{e^{{{- j2\pi km} \mathord{\left/
 {\vphantom {{- j2\pi km} {M}}} \right.
 \kern-\nulldelimiterspace} {M}}}}}$, from which we can derive the power spectral estimation ${{{P}}_m} = {\left| {{{{X}}_m}} \right|^2}$. By finding the peak value of ${{{P}}_m}$ in time, we can estimate the exact CFO ${\varphi _f\left( {k} \right)}$. The estimated CFO and phase values are transferred to the header SOVA block to compensate for the phase of the previous channel estimation ${\varphi _p\left( {k-1} \right)}$ in the Trellis search.

\subsubsection{Header SOVA}
GMSK signals allow ISI to increase frequency efficiency, and the Viterbi detector, a type of sequence detector, is often used in receivers to minimize performance degradation due to ISI. The performance and complexity of the Viterbi detector are mainly determined by the number of states in the state transition diagram represented by Trellis and the method of calculating branch metrics. Methods to minimize the number of states in Trellis have been proposed in several studies [\ref{GMSK1}], [\ref{GMSK2}]. First, there is a method that considers only the part with large interference between symbols and a decision feedback method that determines a part of the estimated signal using detected past symbols [\ref{GMSK1}]. There is also a method that uses terms with large weights by approximating to a pulse amplitude modulation (PAM) signal [\ref{GMSK1}], [\ref{GMSK2}]. This paper applies the SOVA method that outputs soft decision values based on the Viterbi state transition diagram, which considers only the large interference between symbols. Here, soft decision values containing reliability information of the received signal are provided to improve error correction decoding performance. In addition, considering the characteristics of LR-FHSS frame data, we propose an enhanced SOVA scheme that performs both forward and backward Trellis searches while reflecting the tracking results of phase ${\varphi _p}$, CFO ${\varphi _f}$ and Doppler rate ${\varphi _{dr}}$.

In the proposed framework, the header data containing LR-FHSS transmission information is demodulated with high accuracy by using the proposed SOVA method. Particularly, in the header SOVA block, both forward and backward Trellis searches are performed to cope with rapidly changing channel responses due to frequency hopping in the narrowband of 488 Hz. For two sets of the candidate Doppler rates, the forward and backward Trellis searches are utilized from the middle of the header to the last signal, e.g., here, [58:114] symbols, and from the middle [1:57] symbols of a header to the first signal, respectively. As a result, the header SOVA block outputs the soft bit sequences in the forward and backward Trellis search with each Doppler rate candidate. Finally, the three soft bit sequences with the best path metrics are selected. 

\begin{figure}[t]
     \centering
        \includegraphics[width=0.8\columnwidth]{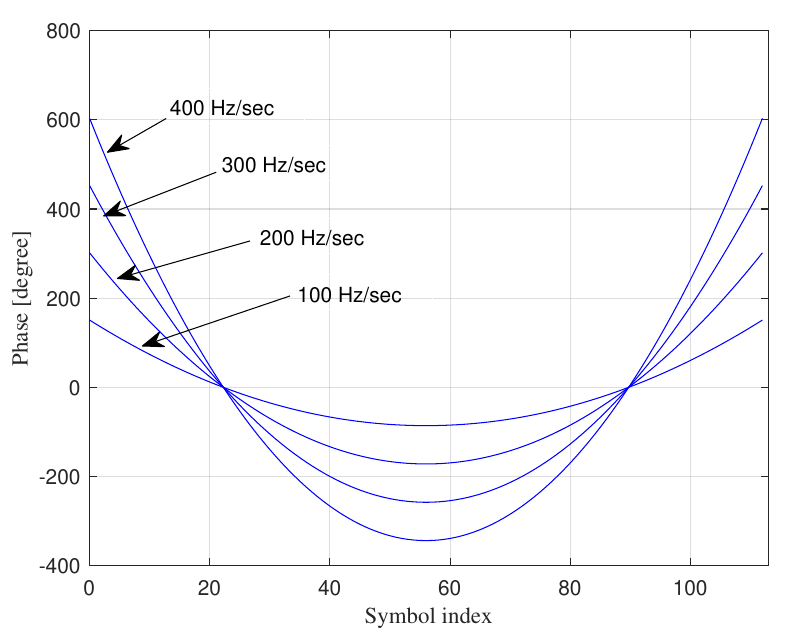}
        \caption{Phase variations of header signal according to the Doppler rate.}
        \label{Phase variation}        
\end{figure}

Fig. 8 shows the phase variations of the header signal according to the Doppler rates of $\{50, 100, 200, 300, 400\}$ Hz/sec. A phase shift of about ${900^\circ}$ is required at a Doppler rate of 400 Hz/sec to make the frequency error zero at the midpoint of the header of 114 symbols. If the Doppler rate is estimated by using the syncword in the middle position of the header, the phase change of the syncword is not large at the Doppler rate of 50 Hz/sec. That is, it is difficult to improve the accuracy of the Doppler rate estimation using syncword. Therefore, the header SOVA scheme is applied with the comparison of the expected Doppler rate candidates, where the header SOVA performs the Trellis search with each Doppler rate candidate. The estimation accuracy of each Doppler rate candidate value is determined by using the bit error rate (BER) after all header bits are demodulated. We consider 11 candidate Doppler rates of $\{0, \pm 80, \pm160, \pm240, \pm320, \pm400\}$ Hz/sec.

\begin{figure}
\centering
\begin{subfigure}{0.46\textwidth}
    \includegraphics[width=\textwidth]{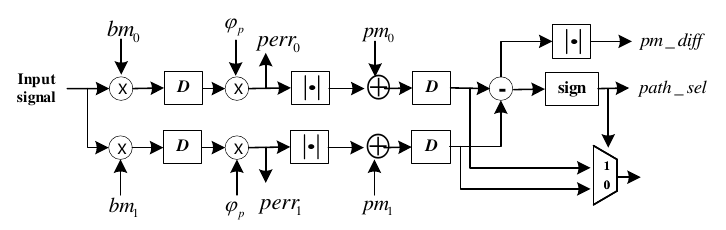}
    \caption{}
\end{subfigure}
\hfill
\begin{subfigure}{0.46\textwidth}
    \includegraphics[width=\textwidth]{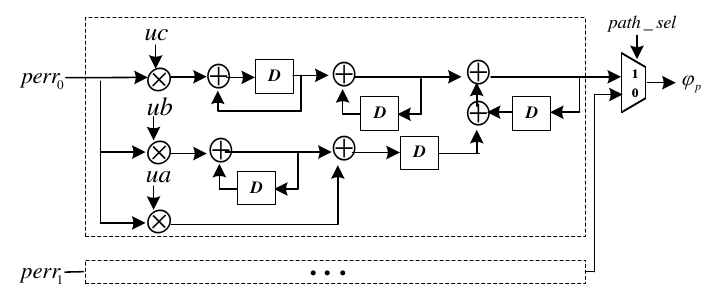}
    \caption{}
\end{subfigure}
\caption{Structure for calculation of (a) the path metric and (b) the channel response in the header SOVA block.}
\end{figure}

For each state in the SOVA scheme, the path selection is performed based on the branch metric $bm_i$, path metric $pm_i$ and phase of the previous channel estimation ${\varphi _{p}\left( {k-1} \right)}$ for $i\in\{0,1\}$, as shown in Fig. 9(a). First, the path metrics are calculated by sequentially compensating the branch metric and the phase of the channel estimation to the input signal. From the two paths connected by each Trellis state, the path with the smaller newly calculated path metric is selected in the path selection $path\_{sel}$. The difference between the two path metrics $pm\_{diff}$ is used for soft bit calculation. To track the phase ${\varphi _p}$, CFO ${\varphi _f}$ and Doppler rate ${\varphi _{dr}}$, the phase error of the signal $perr_i$, for $i\in\{0,1\}$, is transferred to the calculator of the channel estimation, as shown in Fig. 9(b). The $perr_i$ is used as an input for the tracking loop, where $ua$, $ub$ and $uc$ are the coefficients for the phase tracking of the channel estimation, CFO tracking, and Doppler rate tracking, respectively. By these definitions, we can track the phase ${\varphi _p}$, CFO ${\varphi _f}$ and Doppler rate ${\varphi _{dr}}$ as,
\begin{align}
\begin{array}{l}
{\varphi _{dr}}\left( k \right) = {\varphi _{dr}}\left( {k - 1} \right) + uc \cdot per{r_i}, \\
{\varphi _{ds}}\left( k \right) = \sum\nolimits_k {{\varphi _{dr}}\left( k \right)},  \\
{\varphi _f}\left( k \right) = {\varphi _f}\left( {k - 1} \right) + ub \cdot per{r_i}, \\
{\varphi _p}\left( k \right) = {\varphi _p}\left( {k - 1} \right) + {\varphi _{ds}}\left( k \right) + {\varphi _f}\left( k \right) + ua \cdot per{r_i},
\end{array}
\end{align}
where ${\varphi _{ds}}$ is the Doppler shift: the cumulative value of the Doppler rate. In the Trellis search, the bit value of the selected state is stored as the most recent hard bit in the survival path when a new path metric is selected. The $pm\_diff$ is stored as the most recent soft bit in the survival path of the selected path. The hard and soft information for the previous 32 bits are stored and updated with 32 length of the survival path. If more than 32 signals are the input, the oldest soft bit value of the state with the smallest path metric among the 4 states is chosen as the output.

Fig. 10 shows the distribution of the reception errors when using the forward Trellis search and both forward and backward Trellis searches. The error distribution is measured when the payload of 100 bytes is sent 100 times. An issue with the forward Trellis search method is that most errors are concentrated in the front part of the frame, as shown in Fig. 10(a). As shown in Fig. 10(b), both forward and backward Trellis searches distribute the errors to the entire frame, which reduces the PER through the error correction code.

\begin{figure}
\centering
\begin{subfigure}{0.24\textwidth}
    \includegraphics[width=\textwidth]{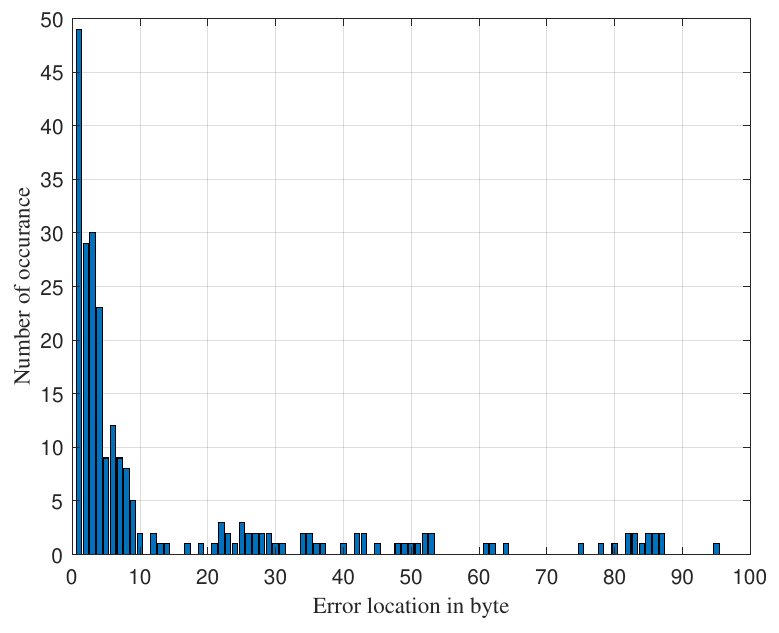}
    \caption{}
\end{subfigure}
\hfill
\begin{subfigure}{0.24\textwidth}
    \includegraphics[width=\textwidth]{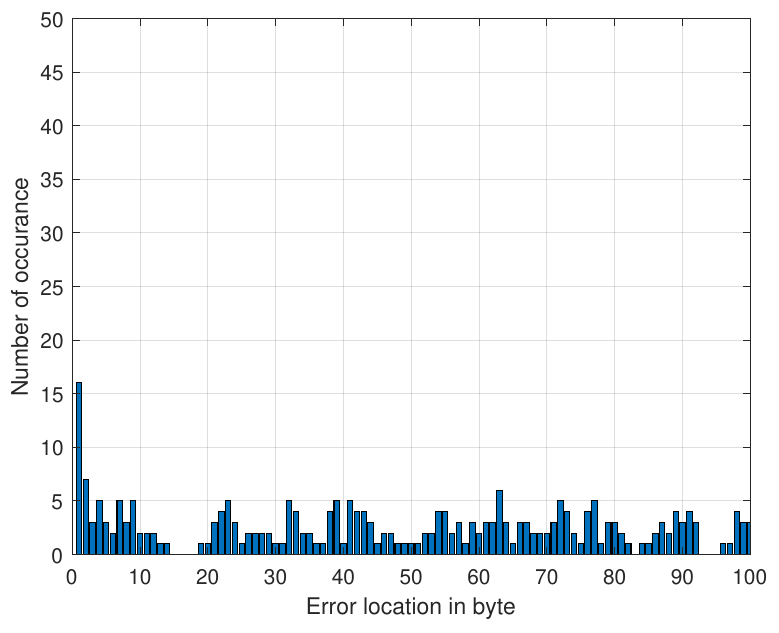}
    \caption{}
\end{subfigure}
\caption{Distribution of the reception error in (a) forward Trellis search and (b) both forward and backward Trellis searches.}
\end{figure}

\subsubsection{Deinterleaver}
The interleaver distributes bits within the length of the header bits to effectively correct bit errors caused by burst noise in the receiver. When reading three soft bit sequences from the header SOVA block, the soft bit sequences are read in the deinterleaving order to recover the original transmission bit order as follows:
\{1, 18, 26, 34, 42, 50, 58, 66, 73, 2, 10, 27, 35, 43, 51, 59, 67, 74, 3, 11, 19, 36, 44, 52, 60, 68, 75, 4, 12, 20, 28, 45, 53, 61, 69, 76, 5, 13, 21, 29, 37, 54, 62, 70, 77, 6, 14, 22, 30, 38, 46, 63, 71, 78, 7, 15, 23, 31, 39, 47, 55, 72, 79, 8, 16, 24, 32, 40, 48, 56, 64, 80, 9, 17, 25, 33, 41, 49, 57, 65\}.

\subsubsection{Viterbi Decoder}
The Viterbi decoder for header decoding with the code rate 1/2 has 64 states, and the operation in each state is performed, as shown in Fig. 11. The $llr\_a$ and $llr\_b$ are the soft bits indicating the reliability of encoded bits, the $s_0$ and $s_1$ are previous states transitioning to the current state $s$, the $a_0$ and $a_1$ are two coded bits output when transitioning from state $s_0$ to state $s$, and the $b_0$ and $b_1$ are two coded bits output when transitioning from state $s_1$ to state $s$. The $pm[s_0]$, $pm[s_1]$, and $pm[s]$ are path metrics in states $s_0$, $s_1$, and $s$, respectively, and the process of finding the state transition path that maximizes path metrics is the operation of the Viterbi decoder. Among the two paths transitioned from states $s_0$ and $s_1$, the state transition path with the larger path metric is selected, and the data bit corresponding to the selected path is added as a new bit to the survival path register in state $s$. After the data bits of 40 length are input twice for demodulation of the encoding bits, the final state information is obtained with the oldest data 6 bits of the survival path register stored in the state with the largest path metric among the 64 states.

\begin{figure}[t]
\centering
    \includegraphics[width=0.9\columnwidth]{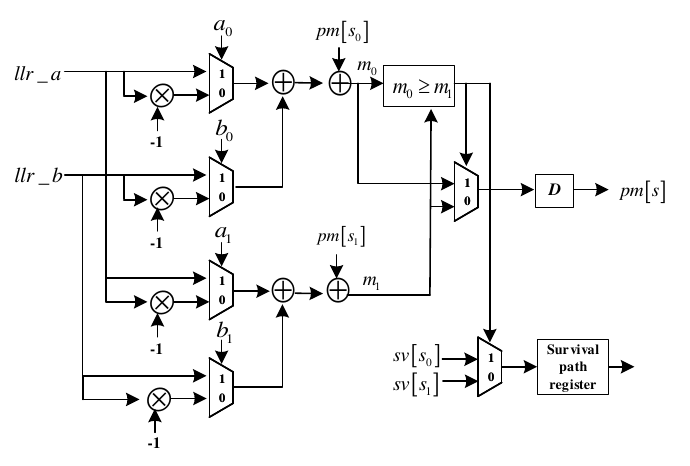}
    \caption{Structure of the Viterbi decoder for header decoding.}
\end{figure}

\subsubsection{CRC8}
The CRC8 block calculates the cyclic redundancy check (CRC) to detect the bit errors of the received header data. If the register value becomes 0 after all header bits, including the CRC of 8 bit, are entered in the CRC8 block, it is determined that there is no bit error.

\subsubsection{Header Decoder}
After the CRC8 check, the header decoder can obtain the payload information necessary for receiving the payload from the header data received without the error. This payload information is transferred to external memory. From the payload information, we can find the location of the entire payload stored in the external memory.

\subsection{Payload Receiver}
In the payload receiver, the payload data stored in the payload buffer passes through the blocks for synchronization and payload decoding. For synchronization of the payload data, the estimation results, such as the Doppler rate, CFO and STO, obtained in the header receiver are reused to synchronize the payload.

\subsubsection{Payload Buffer}
In the payload buffer, the payload data to be demodulated is read with the payload information from the external memory, and is transferred to the synchronization block in units of hopping blocks. The payload information, which contains the channel/hopping sequence in the header, is used to output the payload data in units of hopping blocks. In the synchronization block, the estimation results, such as the Doppler rate, CFO and STO, obtained in the header receiver are reused to synchronize the payload. In addition, the phase rotate, CFO, and phase estimation blocks are performed the same as in the header receiver for the payload SOVA.

\subsubsection{Payload SOVA}
The input of the payload SOVA block is the synchronized payload data in hopping block units. To obtain the soft bits from the GMSK-based payload fragments, the Trellis search with 4 states is applied for the payload demodulation. The payload SOVA has the same role as the header SOVA. However, the length of the input signal is 50 symbols for the payload fragment, unlike the 114 symbols for the header fragment.

\subsubsection{Deinterleaver}
In the LR-FHSS transmitter, the input soft bits of the payload are sequentially increased by 48, and a total of $n$ blocks are stored in memory. Conversely, the output soft bit is read from memory by adding 48 from the initial address 0. For deinterleaving, the data is read in the order of input soft bits, and is repeated until all soft bits of $n$ blocks are read.

\subsubsection{Viterbi Decoder}
The code rates of 1/2, 2/3, and 5/6 are created by puncturing the convolutional code with the code rate of 1/3 as the mother code. The Viterbi decoder performs the decoding process by filling the punctured bits with 0 according to each code rate, that is, depuncturing the bits. The Viterbi decoder for the payload works the same as the Viterbi decoder for the header. 

\subsubsection{CRC16}
To detect the payload bit errors, the CRC calculation is performed identically to the CRC16 of the transmitter. The input bits are collected into bytes in most significant bit (MSB) order and fed into the same circuitry as the CRC16 of the transmitter. After calculating all of the payload data, the value of the shift register is compared with the received 16 bit CRC parity bit. Then, if it is the same as CRC16 parity bit, it is determined to be received without the error.

\begin{table}[t]
    \caption{Simulation Parameters}\label{Table3}
    \centering
    \begin{tabular}{|p{5.8cm}|p{2.1cm}|}
     \hline
     Parameters & Values \\
     \hline
     Carrier frequency $f_c$ &  940 MHz \\
     Number of header replicas $N_H$ & \{3, 2\} \\
     Number of payload fragments $N_F$ & \{18, 9\} \\
     Header duration $T_H$ & 233.472 ms \\   
     Payload duration $T_F$ &  102.4 ms \\      
     OCW &  39 kHz \\
     Grid &  3.9 kHz \\
     Maximum payload size $L$ &  32 bytes \\   
     Coding rate $r$ &  \{1/3, 2/3\} \\
     Number of channels for hopping $N_{CF}$ &  80 \\
     Number of channels for hopping per ED $N_{CF/ED}$ &  10 \\
     Initial STO & \{0, 1/8, 1/4\} \\        
     SFO & \{0 ${\sim}$ 80\} ppm \\      
     CFO & \{0 ${\sim}$ 5/6\} \\ 
     Doppler rate & \{0 ${\sim}$ 400\} Hz/sec  \\
     CCI & \{0 ${\sim}$ 60\} $\%$  \\ 
     \hline
    \end{tabular}
\end{table}
 
\subsection{Simulation Results}
To validate the proposed LR-FHSS transceiver algorithm, the PER performance of LR-FHSS transmission was evaluated under actual LEO satellite channel effects. The simulation parameters are summarized in Table III [\ref{COMML}], [\ref{Previous5}]. The LEO satellite channel effects were considered as follows: the values for the initial STO as \{0, 1/8, 1/4\}, for SFO as \{0, 20, 60, 80\} ppm, for CFO as \{0, 1/6, 3/6, 5/6\}, for Doppler rate as \{0, 320, 400\} Hz/sec, and for CCI as \{0, 4, 12, 20, 40, 60\}\%. The PER performance of the header with $r=1/2$ and payload with $r=1/3$ and $2/3$ was measured under the AWGN environment, serving as a reference performance for comparing the performance degradation effects caused by the LEO satellite channel model.

\begin{figure*}
\centering
\begin{subfigure}{0.45\textwidth}
    \includegraphics[width=\textwidth]{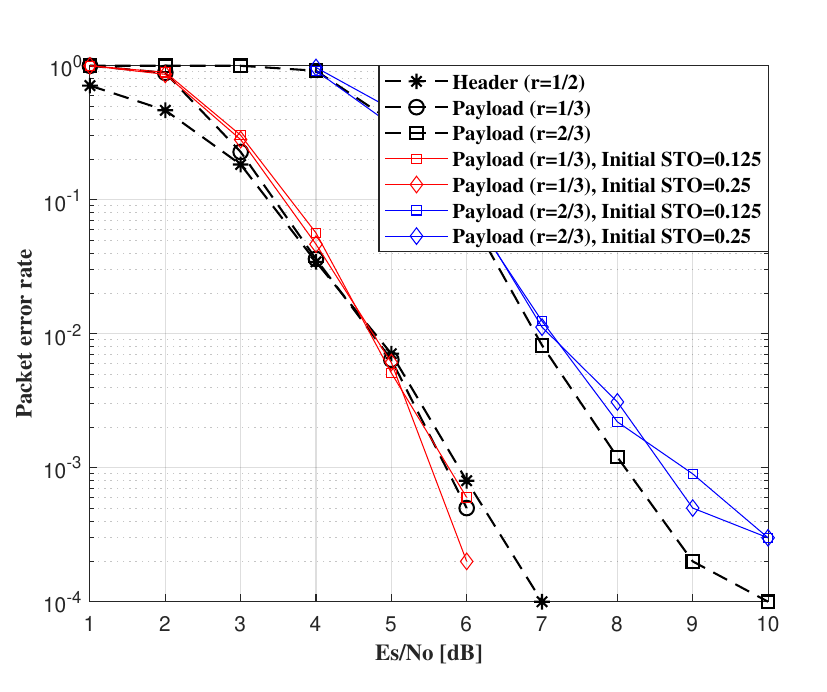}
    \caption{}
\end{subfigure}
\begin{subfigure}{0.45\textwidth}
    \includegraphics[width=\textwidth]{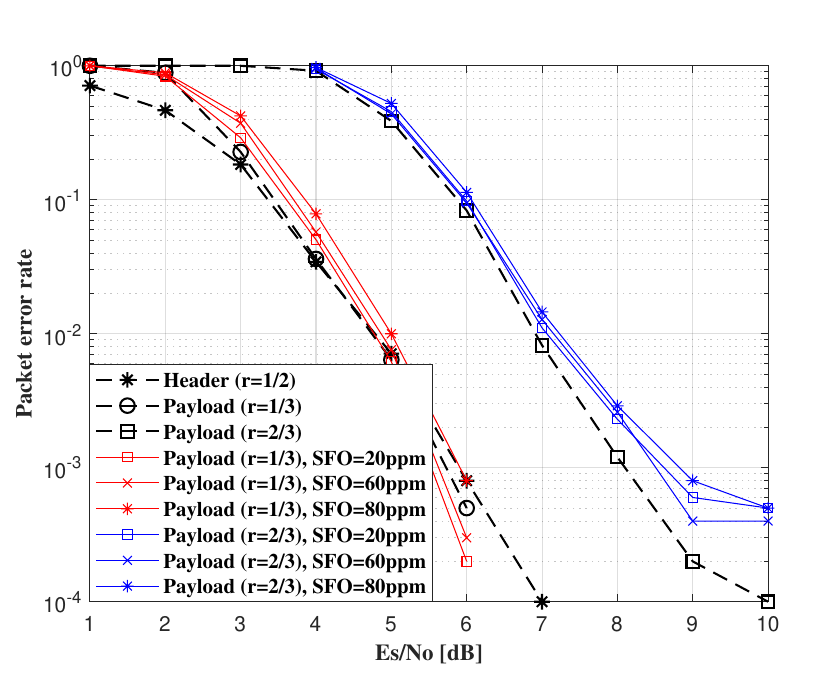}
    \caption{}
\end{subfigure}
\caption{Packet error rate of the LR-FHSS signal according to the (a) initial STO and (b) SFO.}
\end{figure*}

\begin{figure*}
\centering
\begin{subfigure}{0.45\textwidth}
    \includegraphics[width=\textwidth]{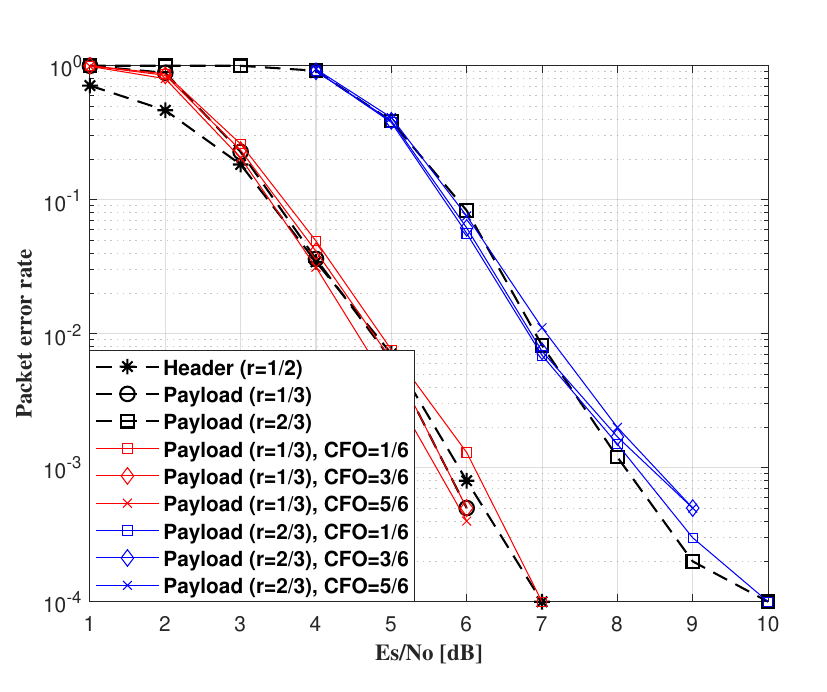}
    \caption{}
\end{subfigure}
\begin{subfigure}{0.45\textwidth}
    \includegraphics[width=\textwidth]{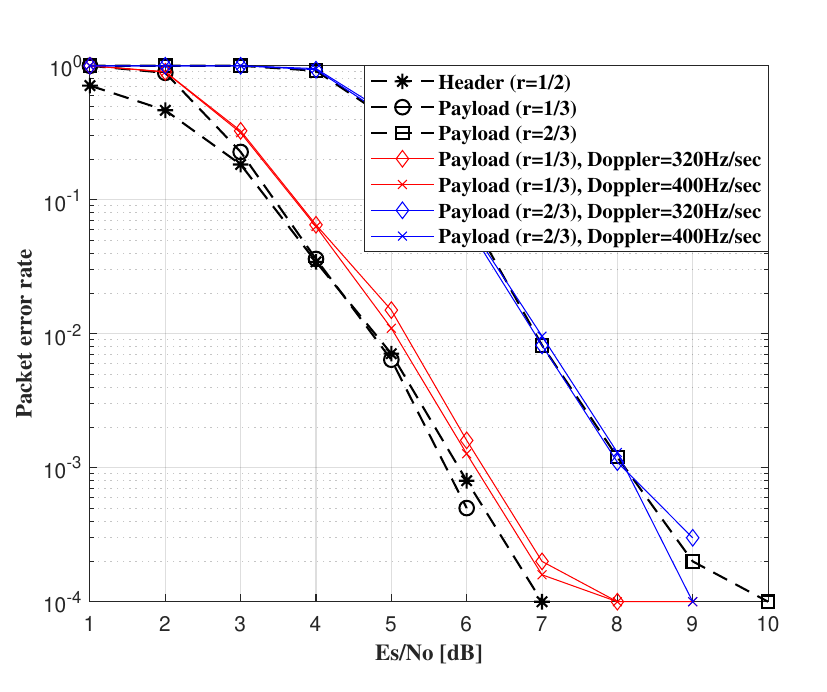}
    \caption{}
\end{subfigure}
\caption{Packet error rate of the LR-FHSS signal according to the (a) CFO and (b) Doppler rate.}
\end{figure*}

\begin{figure}[t]
\centering
    \includegraphics[width=0.45\textwidth]{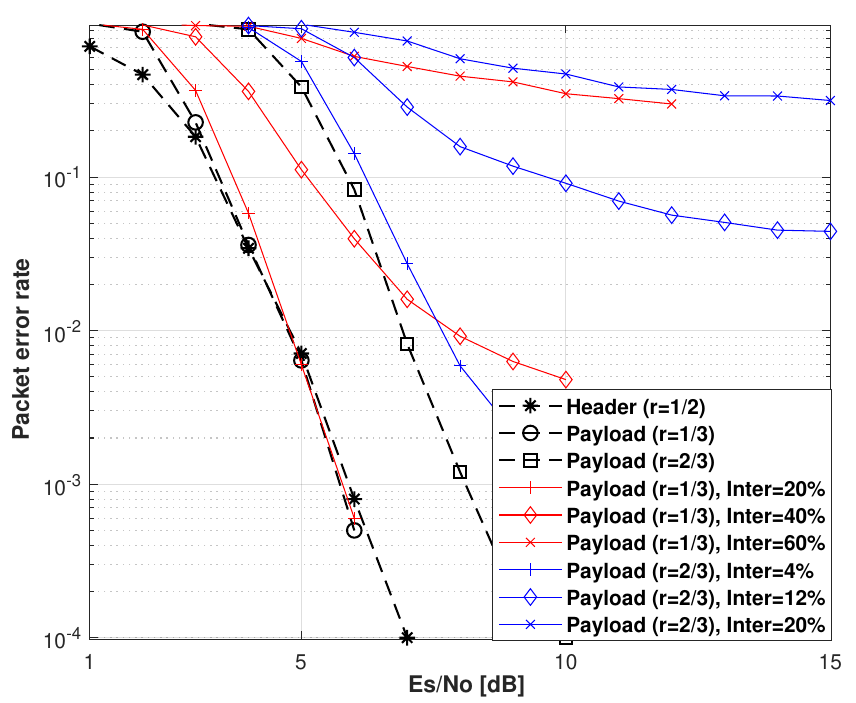}
    \caption{Packet error rate of the LR-FHSS signal according to the CCI.}
\end{figure}

Fig. 12 shows the PER performance of the LR-FHSS signal according to the initial STO and SFO, which occurs due to the long-term instabilities of the receiver oscillator. As shown in Fig. 12(a), the simulation was performed at $r=1/3$ and $2/3$, and the PER performance at the initial STO of 0 was identical to that in the AWGN environment. According to the initial STO, the PERs at $r=1/3$ make little performance difference, while the performance difference of 1 dB at a PER of $10^{-3}$ is observed at $r=2/3$ due to the low coding gain. Fig. 12(b) shows the PER performance of the LR-FHSS signal for the different SFOs. As SFO increases, the PER gradually rises due to the increased timing errors in the last symbol of the packet. The performance of the proposed symbol timing estimation block is verified, showing only a slight gap of 0.5 dB compared to the reference performance.

Fig. 13(a) illustrates the PER performance of the LR-FHSS signal under different CFOs. There is no difference between the performances according to CFO and the reference performance, indicating that the CFO compensation block consisting of coarse estimation, fine estimation, and tracking blocks nearly compensates the CFO. Fig. 13(b) shows the PER performance of the LR-FHSS signal with different Doppler rates. At $r=2/3$, there is little performance difference according to the Doppler rates, while, at $r=1/3$, the performance difference occurs about 0.5 dB. The performance degradation at $r=1/3$ may be caused by miss detection in the signal detector because the signal detector of the GMSK-based LR-FHSS are affected by the Doppler rate.

Fig. 14 illustrates the PER performance of the LR-FHSS signal as affected by CCI. The simulation was performed by inputting the number of hopping blocks in which the CCI occurs as a percentage. At $r=1/3$ and $2/3$, the CCIs of \{20, 40, 60\}\% and \{4, 12, 20\}\% were applied, respectively. Since the impact of the CCI on PER performance is different depending on the coding rate, the different interference percentages were applied for each code rate. As a result, for $r=1/3$, the error floor starts at Es/No of 8 dB under CCI conditions of 40\% or more, while, for $r=2/3$, the error floor starts at Es/No of 8 dB under the CCI conditions of 12\% or more. This demonstrates that strong coding gain can withstand higher CCI.

\section{System-level Implementation and Verification via Laboratory Test}\label{sec: Section4}
Using the proposed LR-FHSS transceiver design including the synchronization algorithm, we implemented the LR-FHSS transmitter testbed based on the ASIC chipset and the receiver testbed based on the FPGA, as shown in Fig. 15. To achieve the target QoS of the DtS-IoT use cases, i.e., environmental monitoring that requires data rates of less than 1 kbps, quite relaxed latency, and low power conditions, as shown in Table I, the ASIC chipset process was specially adopted for low-power consumption and miniaturization of the terminal. The implementation of the LR-FHSS transceiver was verified via laboratory tests using multiple terminal transmissions over wired and wireless communications. Finally, the proposed DtS-IoT system including the LR-FHSS transceiver was constructed and verified through laboratory tests such as a system operation request and response.

\begin{figure}
\begin{subfigure}{\columnwidth}
    \centering
    \includegraphics[width=0.9\columnwidth]{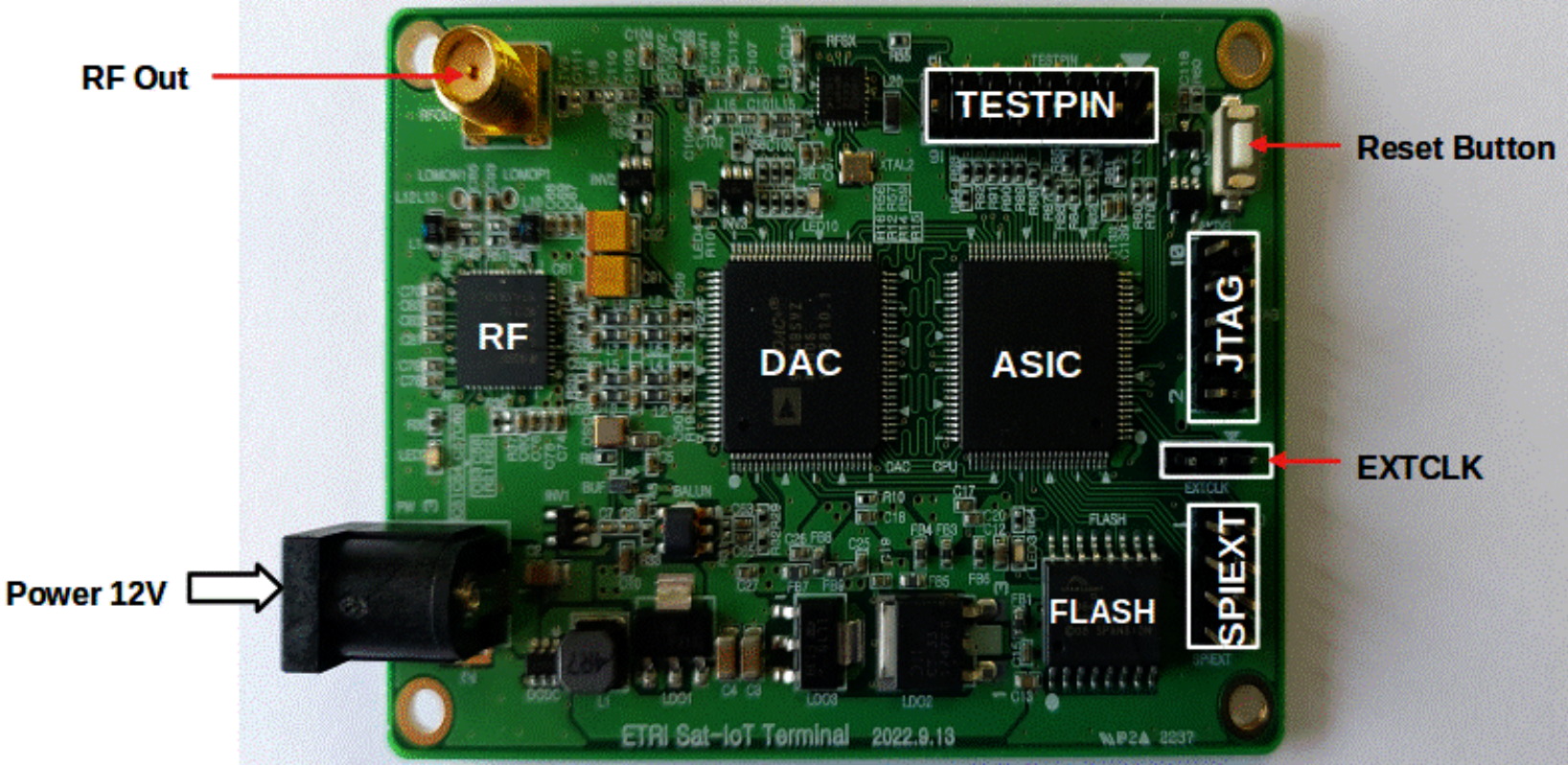}
    \caption{}
\end{subfigure}
\begin{subfigure}{\columnwidth}
    \centering
    \includegraphics[width=0.8\columnwidth]{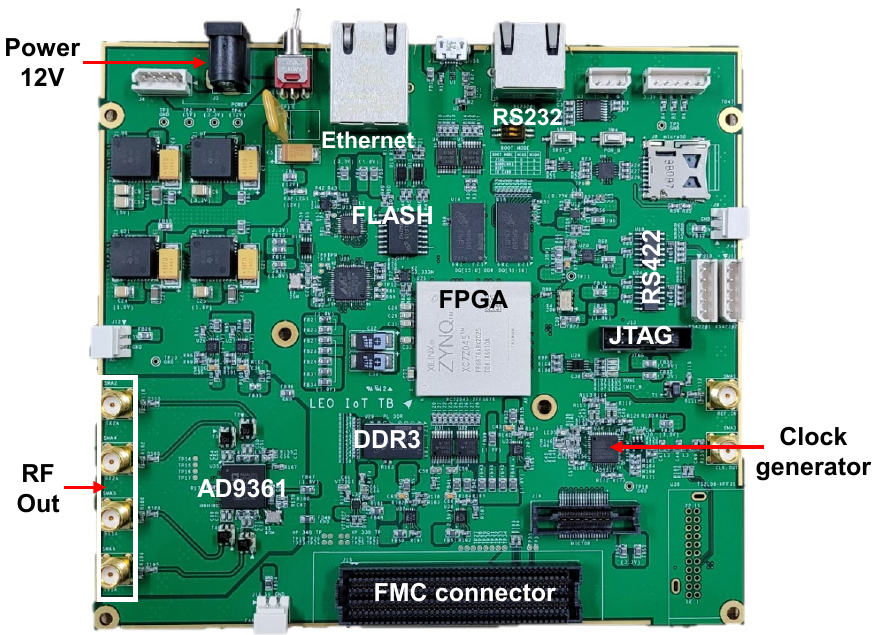}
    \caption{}
\end{subfigure}
\caption{(a) Transmitter testbed based on the ASIC chipset and (b) receiver testbed based on FPGA for LR-FHSS transmission.}
\end{figure}

\subsection{Implementation of ASIC-based Transmitter and FPGA-based Receiver Testbeds}
The ASIC chipset for LR-FHSS transmition was built via the system on chip (SoC) process with 0.18 um and 0.6 Mgate by the foundry company SMIC. Fig. 15(a) shows the LR-FHSS transmitter testbed based on the ASIC chipset, which has the following features:
\begin{itemize}
\item CPU: 32-bit RISC-V (RV32I instruction set)
\item Memory: RAM for chip program/data memory (32 Kbytes), ROM for chip boot memory (4 Kbytes)
\item Flash controller: Serial interface (external device of S25FL064P with 64 Mbits and 3.0 V)
\item Timer: 16-bit timer with external input, watch-dog with reset \& interrupt support, real time clock
\item Interface: UART (8 bits), SPI (128 bits), I2C (8 bits)
\item GPIO: General purpose port 0 (16 bits)
\item Power save mode: Deep sleep, sleep, idle
\item Package: 100LQFP (8mA per pin)
\item Die area: 621 um x 265 um
\end{itemize}
In the ASIC chipset, the internal central processing unit (CPU) reads the external sensor information through the sensor interfaces, such as the universal asynchronous receiver transmitter (UART), serial peripheral interface (SPI) and inter-integrated circuit (I2C), and stores it in the internal memory. Next, the transmission function is performed using the LR-FHSS waveform. The internal CPU of RISC-V supports the 32-bit base integer instruction set among the RV32I instruction sets. The read only memory (ROM) where the boot loader is stored and performs the function of copying the user program stored in flash to the random access memory (RAM) when the chip boots. The RAM is a space that stores user programs, and the CPU reads user programs and performs desired operations. The external flash is controlled through the quad SPI (QSPI) flash module, and the port 0 is composed of 16 bits and performs general-purpose input/output (GPIO) functions. Based on the ASIC chipset with the above features, we configured the LR-FHSS transmitter testbed including the AD9776 digital-to-analog converter (DAC), ADRF6755 modulator, flash memory and interfaces. Here, the DAC and RF modulator chips are controlled from an external CPU through the SPI interface.

Fig. 15(b) depicts the LR-FHSS receiver testbed based on FPGA. Given the constraints of terminals in terms of power consumption and size, a receiver intended for placement within the LEO satellite payload necessitates regenerative or onboard processing capabilities, for which FPGA-based implementation is suitable. The FPGA-based receiver testbed, utilizing the Xilinx Zynq-7000 SoC, incorporates components such as the AD9361 RF tuner, AD9517 clock generator, flash and DDR3 memory. The AD9361 RF tuner, equipped with integrated 12-bit DACs and analog-to-digital converters (ADCs), facilitates transmission and reception functions. DDR3 SDRAM serves as external memory for the classification and storage of up to $N_{CF} =$3120 LR-FHSS signals channelized in the signal detector of the LR-FHSS receiver. Various interfaces including RS232, RS422, FPGA mezzanine card (FMC) connector, mictor connector, ethernet, and JTAG are utilized.

\begin{figure*}[t]
\centering
    \includegraphics[width=0.9\textwidth]{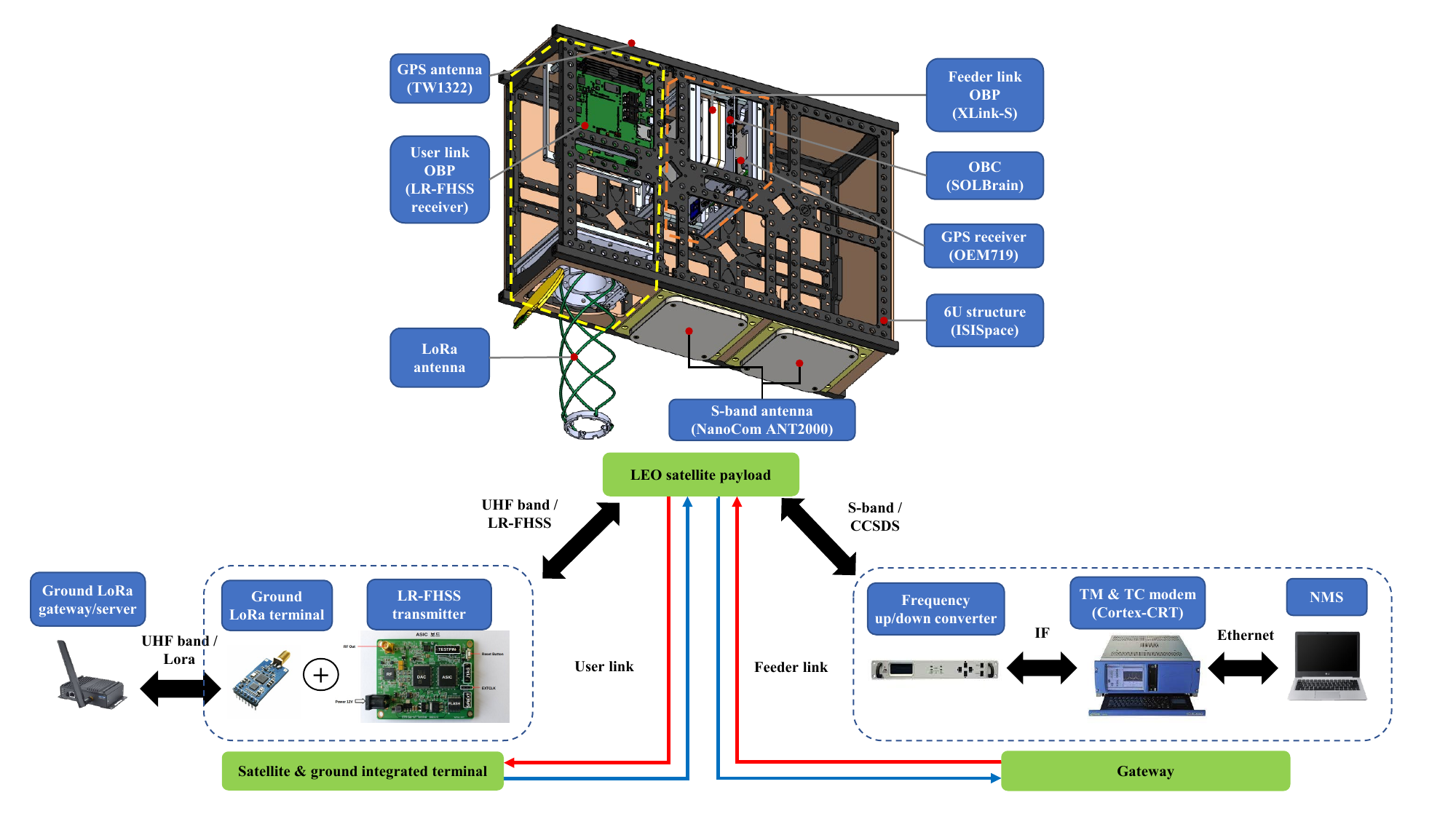}
    \caption{Test diagram for verification of the entire DtS-IoT system.}
\end{figure*}

\subsection{Test diagram for verification}
To verify the overall DtS-IoT system architecture suggested in Fig. 1, we built a test diagram, as shown in Fig. 16. This verification was carried out on the ground by establishing the DtS-IoT communication system by the development of the LEO satellite payload with an engineering model (EM) level. For the user link between the satellite and ground integrated terminals and the LEO satellite payload, the LR-FHSS transmitter sends the collected sensor data to the user link OBP of the LR-FHSS receiver in UHF band frequency, i.e., 940 MHz, supported by our self-developed LoRa antenna. In addition, the satellite and ground integrated terminal includes the module for the LoRa transmission and reception to link the ground gateway and server.

In the user link OBP, the collected sensor data received from the terminals is demodulated and stored temporarily in the DDR3 memory. The demodulated data from the multiple terminals is delivered to OBC, here SOLBrain, and stored, while the LEO satellite payload flies in the orbit until it can communicate with the gateway. In OBC, the GPS information is received at GPS receiver OEM719 with a GPS antenna of TW1322 which includes the epoch time and the location data, such as the coordinated universal time (UTC), latitude, longitude, and altitude. When the connection to the gateway is established, the stored data is transmitted from the feeder link OBP of the XLink-S to the gateway in S-band frequency with 2.4 GHz, supported by the S-band antenna of NanoCom ANT2000.

For the feeder link between the LEO satellite payload and the gateway, the feeder link OBP transmits stored data to the TM and TC modem of Cortex-CRT via the frequency up/down converter, which converts the S-band frequency to the intermediate frequency (IF) of 70 MHz. The TM and TC modem utilize the consultative committee for space data systems (CCSDS) protocol [\ref{CCSDS}]. This protocol is specifically designed for communication over a space link or within a network comprising one or multiple space links. A space link refers to the communication link between a spacecraft and its associated ground system. In the gateway, the NMS receives the demodulated data from the TM and TC modem via ethernet and processes it in the network protocol module. In addition, the functions for beacon transmission management and satellite and terminal information management can be performed.

\begin{figure}
\begin{subfigure}{\columnwidth}
    \centering
    \includegraphics[width=0.95\columnwidth]{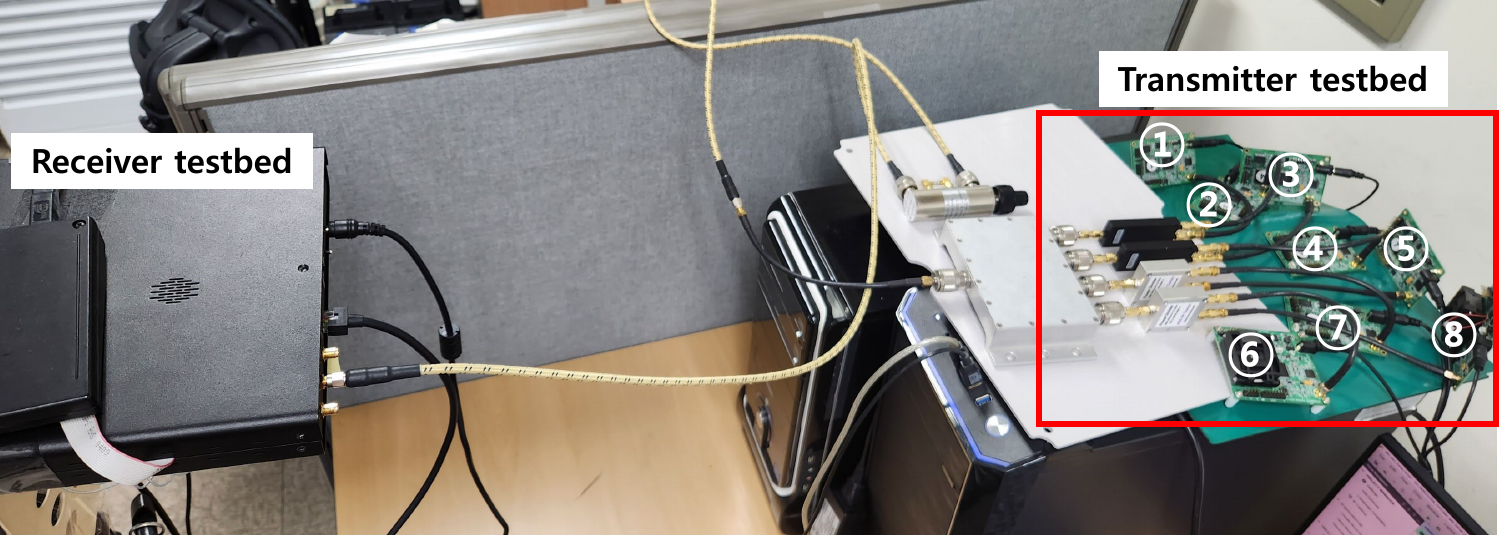}
    \caption{}
\end{subfigure}

\begin{subfigure}{\columnwidth}
    \centering
    \includegraphics[width=0.95\columnwidth]{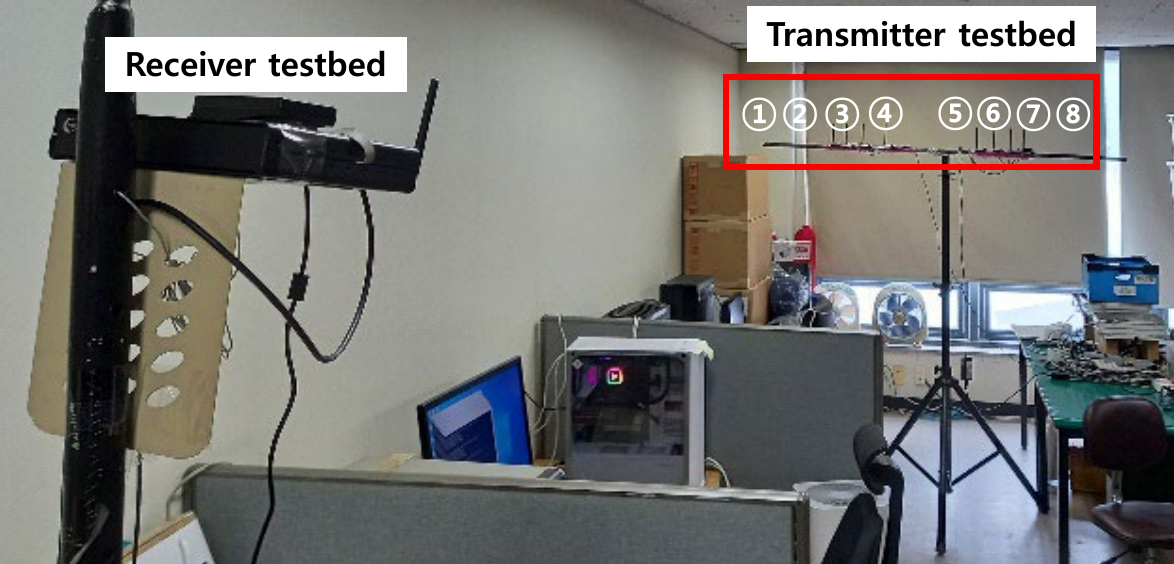}
    \caption{}
\end{subfigure}
\caption{Laboratory test environments for LR-FHSS transmission between the eight transmitters and one receiver over (a) wired communication and (b) wireless communication.}
\end{figure}

\subsection{Laboratory Test Results}
In our laboratory tests, the performance of user link LR-FHSS transmission was analyzed under various conditions with different transmission intervals, code rates and number of headers. The data transmission function of the CCSDS format was verified using a commercial modem of the feeder link. Finally, the entire DtS-IoT system was verified by checking the satellite network access process from the terminal.

\subsubsection{User Link Verification}
Fig. 17(a) and 17(b) illustrate the laboratory test environments for LR-FHSS transmission between the eight ASIC-based transmitter testbeds and single FPGA-based receiver testbed over wired communication and wireless communication, respectively. In the wired communication test, the output of eight transmitters was input to the receiver. In the wireless communication test, four transmitters spaced 8 cm apart were divided into two groups, with an interval of 45 cm between them. The distance between the center of the transmitters and the receiver was 4.7 m. Fig. 18 illustrates the spectral characteristics of the output signal from one of the LR-FHSS transmitters in a wired communication environment. The figure depicts the frequency hopping signals with an occupied bandwidth (OBW) of 488 Hz within an OCW of 1.5 MHz. The spectrum exhibits an intermittent, empty characteristic in the frequency domain during non-hopping intervals.

\begin{figure}[t]
\centering
    \includegraphics[width=0.9\columnwidth]{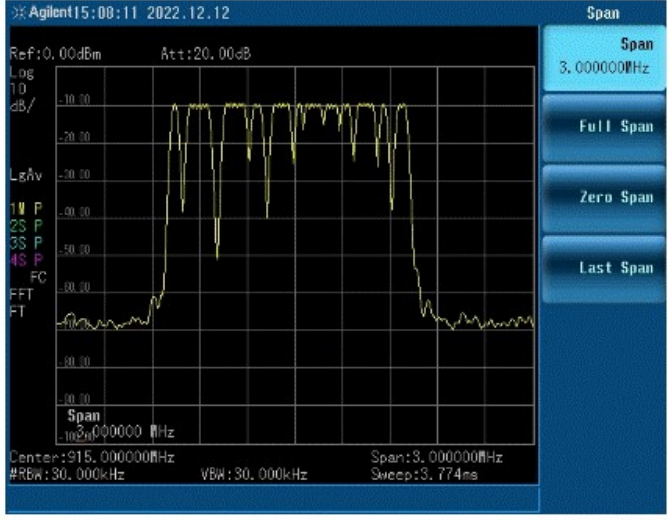}
    \caption{Spectrum of the LR-FHSS transmission.}
\end{figure}

Fig. 19 displays the PER performance of LR-FHSS transmission at various code rates and transmission intervals, corresponding to different numbers of headers, in a laboratory test environment over wired communication. Code rates of \{5/6, 2/3, 1/2, 1/3\} were utilized. The packet emission interval of each transmitter was randomly selected in the short interval from 4 s to 6.56 s and in the long interval from 20 s to 32.8 s. For each transmission, the packet length increased sequentially to [10:10:100] bytes. In this simulation, the actual IoT data transmission-like environment was designed by adopting random transmission intervals and packet lengths. From these parameters, the PER was measured according to the number of headers from 1 to 4, where increasing the number of headers can improve the reception performance with low-spectral efficiency. Via simulations, we found the optimal number of header repetitions between the burden of the header repetition and reception performance. Compared to the PERs of LR-FHSS transmission with a short interval, those with a long interval are observed to be lower due to a lower packet collision rate. In both short and long intervals, as the number of headers increased, the PER decreased because of reliable header detection. At low code rates of 1/2 and 1/3, the PER reduction becomes significant, e.g., at the code rate of 1/3, the PER is slightly worse than at 1/2. This is because at the very low code rate, the number of hopping blocks may increase due to the packet length increased by encoding, leading to an increasing packet collision rate. Consequently, we conclude that less header repetition is effective at lower code rates.

\begin{figure}[t]
\centering
    \includegraphics[width=0.9\columnwidth]{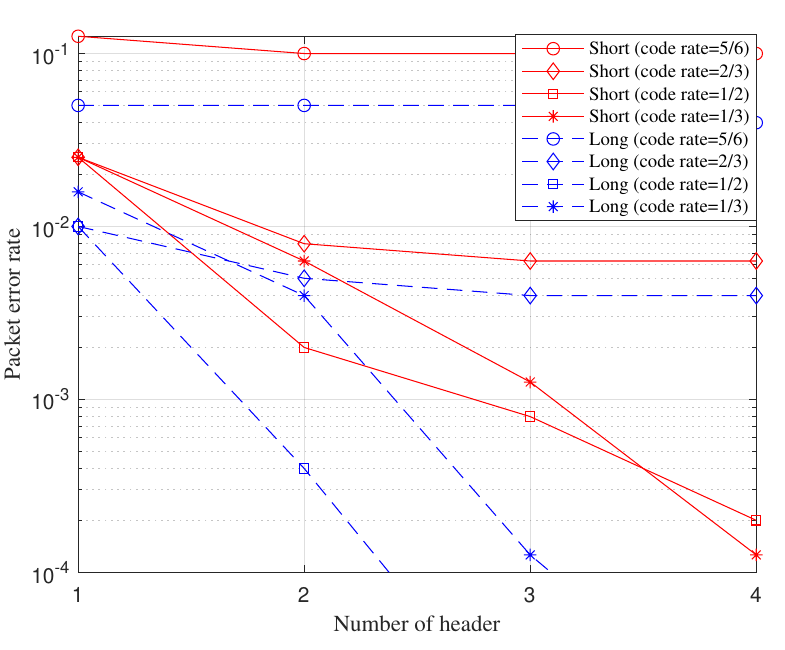}
    \caption{Packet error rate at different code rates and transmission intervals according to the number of headers over wired communication.}
\end{figure}

\begin{figure}[t]
\centering
    \includegraphics[width=0.9\columnwidth]{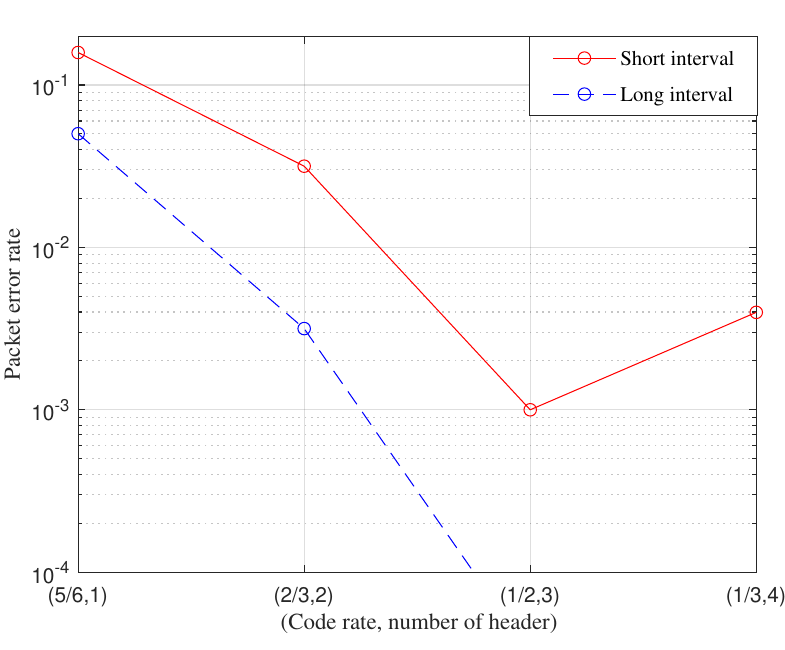}
    \caption{Packet error rate according to the code rate with the number of headers in the short and long intervals over wireless communication.}
\end{figure}

Fig. 20 illustrates the PER performance of LR-FHSS transmission according to the code rate and the number of headers, considering both short and long intervals, in a laboratory test environment over wireless communication. Here, the number of headers for each code rate was adopted as (5/6,1), (2/3,2), (1/2,3), and (1/3,4): combinations defined in the LoRaWAN standard [\ref{Previous1}]. Compared to Fig. 19, to use the wired transmission link, the wireless transmission test was performed in short and long intervals. As the code rate decreases and the number of headers increases, the PER gradually decreases. However, even if four headers are used at the lowest code rate of 1/3, the PER performance of the short interval is higher than the code rate of 1/2 with three headers, which is the previous measurement condition. From these results, we estimate that using a code rate of 1/3 at short intervals can cause frequent collisions, while at long intervals, the use of a stronger code rate with increasing headers reduces the PER. Through our user link transmission test results, we demonstrate that LR-FHSS transmission supports data rates of up to 1523 kHz, which is a KPI for environmental monitoring in low-power-based ASIC terminals.

\subsubsection{Feeder Link Verification}
For the feeder link transmission, as shown in Fig. 16, the S-band frequency is used between the feeder link OBP and the TM and TC modem via the frequency up/down converter, which converts the S-band frequency to the IF of 70 MHz. In addition, the symbol rate of 40 Msps and 128 ksps is used for TM downlink and TC uplink, respectively, with the modulation method of BPSK or QPSK. The data format follows the CCSDS protocol [\ref{CCSDS}] of the inner format and the TM and TC protocol of the outer format. That is, the CCSDS protocol is inserted into the data part of the TM and TC protocol. After the feeder link transmission data following CCSDS and the TM and TC protocols arrives in the LEO satellite payload, the XLink protocol is applied in OBC and OBP modules. For DtS-IoT system verification, Fig. 21 shows the laboratory test environment, which is composed of the gateway, LEO satellite payload and satellite and ground integrated terminal. Here, the gateway includes the NMS, the TM and TC modem, and the frequency up/down converter.

\begin{figure}[t]
\centering
    \includegraphics[width=\columnwidth]{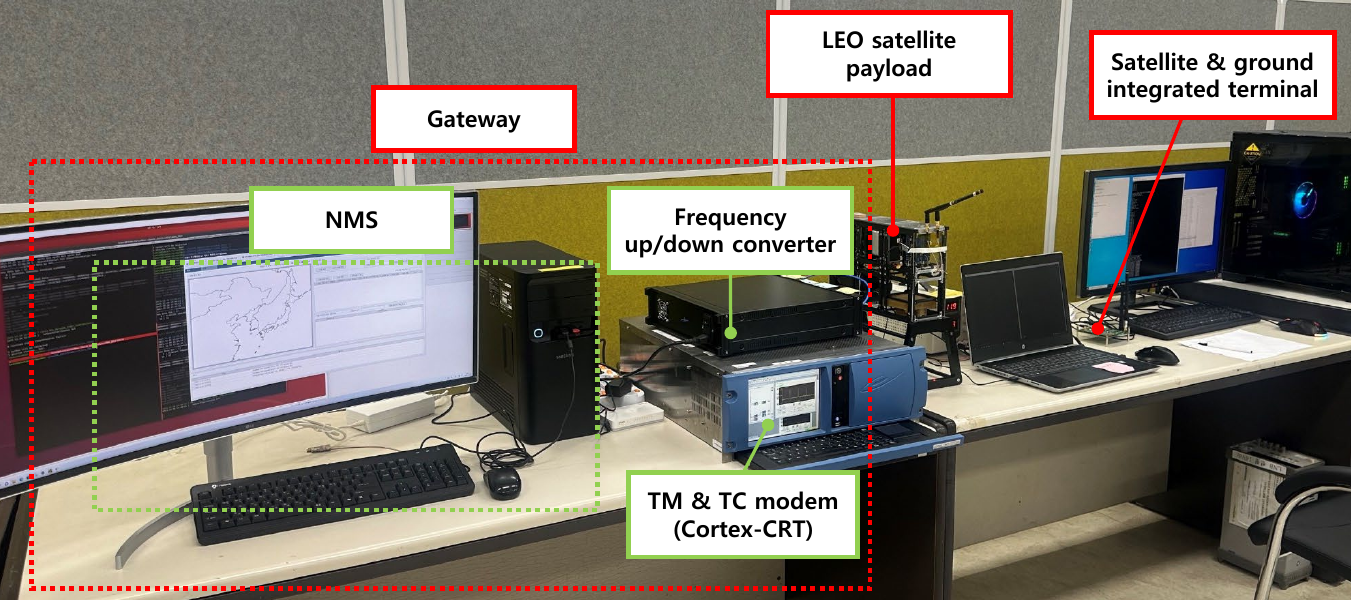}
    \caption{Laboratory test environment for DtS-IoT system verification.}
\end{figure}

\begin{figure}
\centering
\begin{subfigure}{\columnwidth}
    \includegraphics[width=\columnwidth]{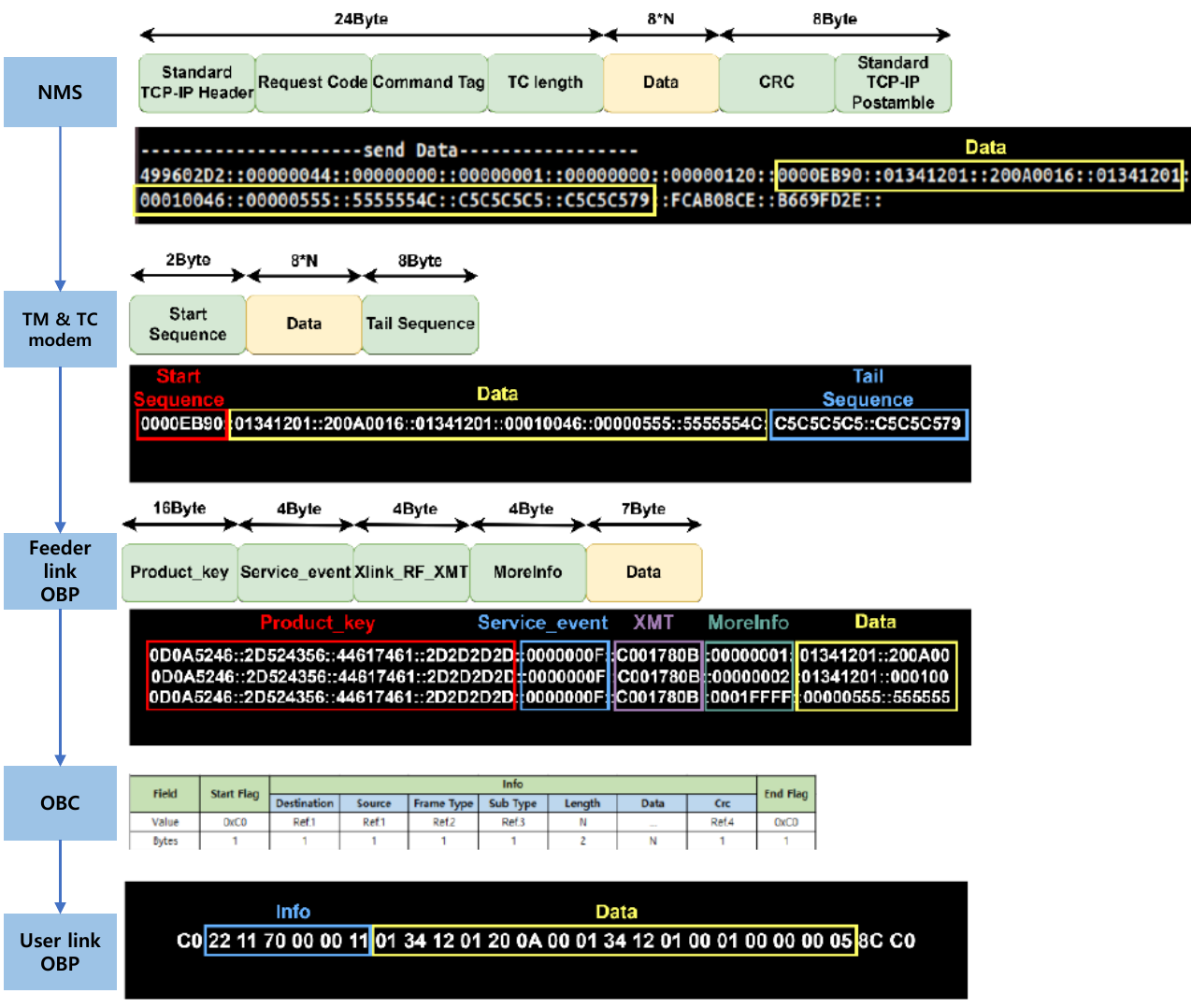}
    \caption{}
\end{subfigure}

\centering
\begin{subfigure}{\columnwidth}
    \includegraphics[width=\columnwidth]{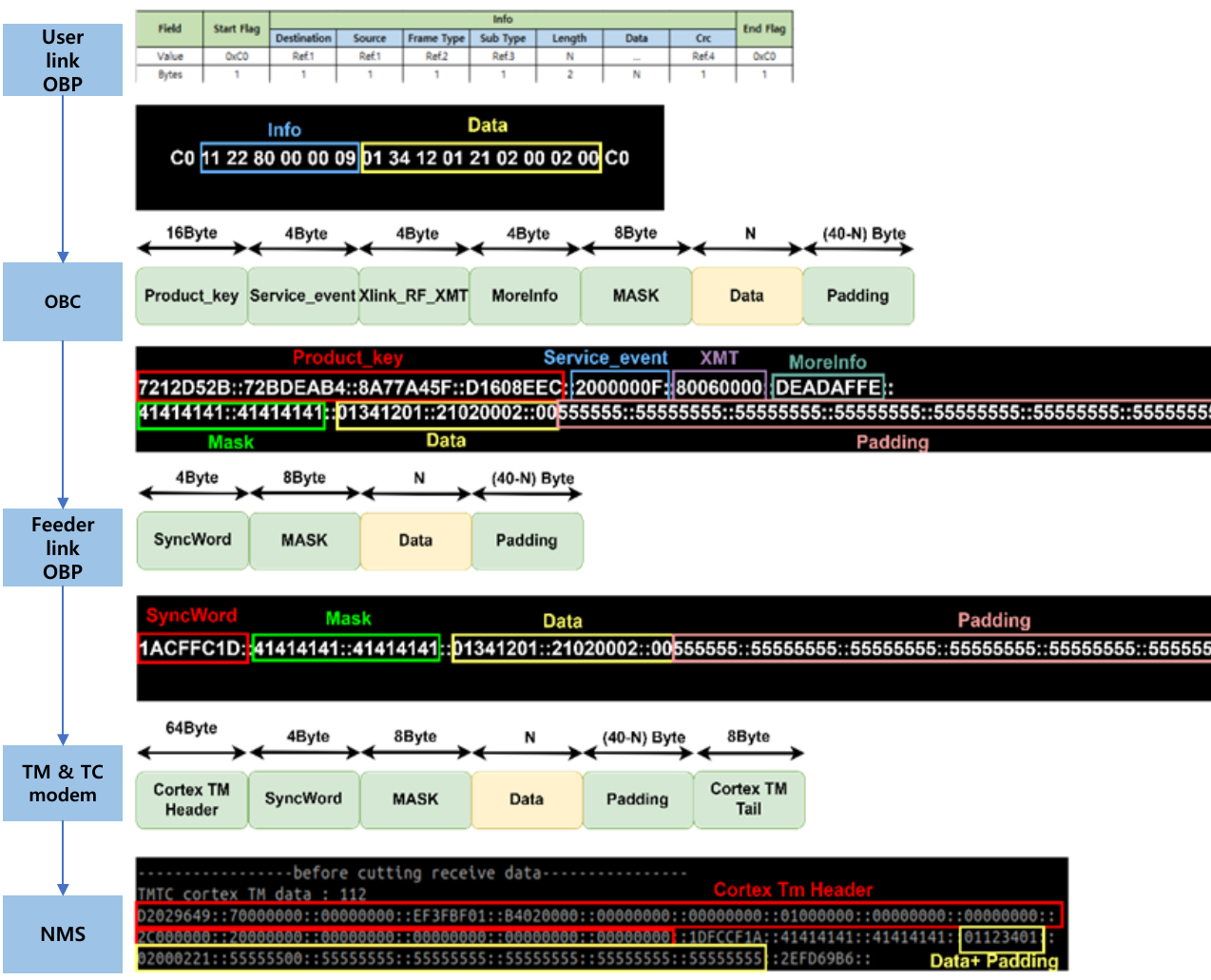}
    \caption{}
\end{subfigure}
\caption{Results of the feeder link transmission verification for (a) TC uplink and (b) TM downlink.}
\end{figure}

Fig. 22 shows the results of the feeder link transmission verification for the TC uplink and TM downlink. As illustrated in Fig. 22(a) for the TC uplink, the NMS module first transmits the TM and TC protocol, which consists of the TCP-IP header, request code, command tag, TC length, data, CRC and TCP-IP postamble. Here, the CCSDS protocol consisting of a start sequence, BCH encoded data, and tail sequence is inserted into the data part of the TM and TC protocol. After the feeder link transmission data following CCSDS and TM and TC protocols arrives in the LEO satellite payload, the XLink protocol is applied in the OBC and OBP modules. The XLink protocol consists of the product key, service event, XMT, and data. The data part of the XLink protocol is included after 1 byte of CRC is excluded from the BCH encoded data of the CCSDS protocol. Finally, the user link OBP is received after the padding bits of 0x55 are excluded. For the TM downlink, the user link OBP first transmits the XLink protocol, which consists of the product key, service event, XMT, moreinfor, mask, and data with the padding, as shown in Fig. 22(b). In the feeder link OBP, the syncword is added into the mask and data with padding of the XLink protocol. This data, generated from the LEO satellite payload, is received at the NMS module by adding the header and tail as it passes through the TM and TC modem.

\subsubsection{System Verification}
Figs. 23(a) and 23(b) measure the throughput performance of the proposed DtS-IoT system with the unslotted and slotted Aloha multiple access schemes, respectively. The performance measurements apply the following parameters: a satellite connection time of 1200 s, number of devices 1500, packet length of 20 bytes, and beacon interval of 120 s. The maximum throughput is theoretically calculated to be 0.18 and 0.37 for the unslotted and slotted Aloha multiple access schemes, respectively [\ref{DTS2}]. In Fig. 23(a), the unslotted Aloha multiple access scheme shows a maximum throughput of 0.12 when using only one channel, and a maximum throughput of 0.38 when using three channels. Basically, more than two channels are required to satisfy the theoretical throughput of 0.18. In Fig. 23(b), the slotted Aloha multiple access scheme shows a maximum throughput of 0.24 when using only one channel, and a maximum throughput of 0.7 when using three channels, meaning more than two channels are required to satisfy the theoretical throughput of 0.37. These results demonstrate that the LR-FHSS transceiver-based DtS-IoT system enables high throughput communication link for future 6G networks.

\begin{figure}
\centering
\begin{subfigure}{0.5\textwidth}
    \includegraphics[width=\textwidth]{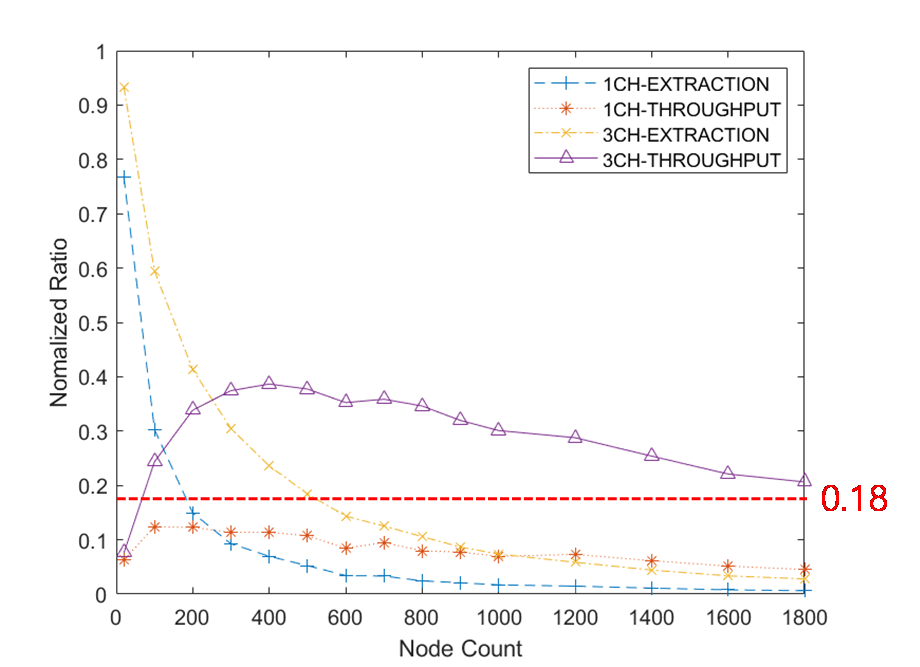}
    \caption{}
\end{subfigure}
\begin{subfigure}{0.5\textwidth}
    \includegraphics[width=\textwidth]{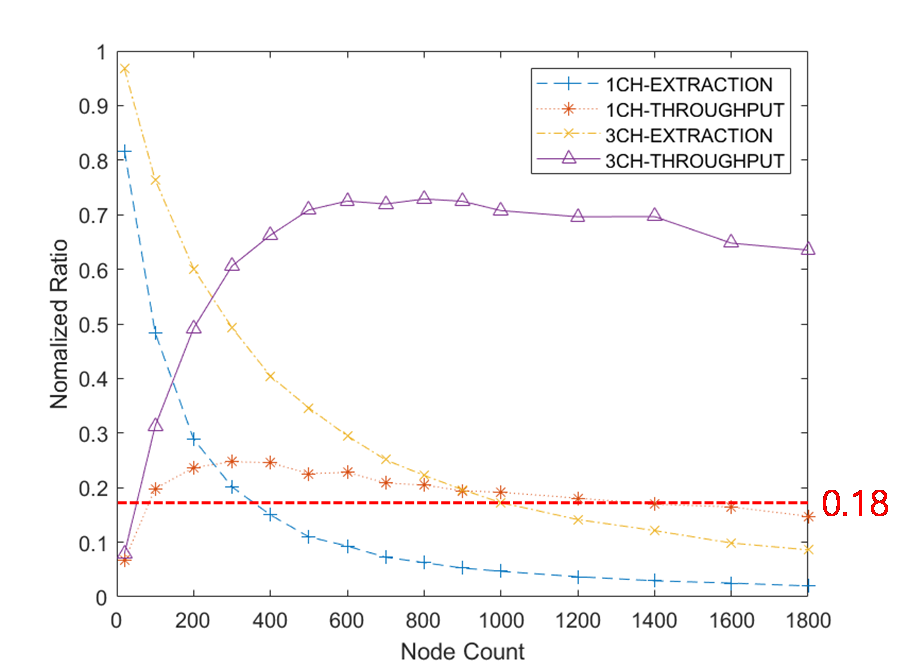}
    \caption{}
\end{subfigure}
\caption{Throughput performance of DtS-IoT system with (a) the unslotted and (b) slotted Aloha multiple access schemes.}
\end{figure}

\section{Conclusions}\label{sec:multiple_orbits}
This paper presented the system architecture and design details for a DtS-IoT system with an LR-FHSS transceiver and its feasibility validation by implementing a testbed and conducting laboratory tests. In this study, we established high-level system architecture, target DtS-IoT use cases and associated KPIs, and provided a design overview of the LR-FHSS transceiver and DtS-IoT system. The LR-FHSS transceiver was designed by using a robust synchronization algorithm based on the GMSK symbol mapping scheme to withstand LEO satellite channel environments. In particular, it includes signal detection for the simultaneous reception of numerous frequency hopping signals and enhanced SOVA for header and payload receptions. To validate the proposed DtS-IoT system, we presented the implementation details for the transceiver testbeds using an ASIC chipset and an FPGA and the entire DtS-IoT system. The DtS-IoT system, mainly composed of the satellite and ground integrated terminal, LEO satellite payload and gateway, was constructed in a laboratory for demonstration. We demonstrated that the entire DtS-IoT system, including to the LR-FHSS transceiver, can perform transmission on the user link and feeder link. Simultaneously with the demonstration of DtS-IoT system functionalities, a link performance assessment to check whether the LR-FHSS communication system can endure LEO satellite channel environments was conducted. The simulation tests showed that an initial STO up to 1/4, SFO up to 80 ppm, CFO up to 5/6 of the symbol rate, Doppler rate up to 400 Hz/sec, and CCI up to 40 percent can be considered. Furthermore, DtS-IoT systems with unslotted and slotted Aloha multiple access schemes were measured to support throughputs of over 0.18 and 0.37, respectively.

This study provides valuable insights into the infinite applicability of DtS-IoT systems and guides readers in implementing an LR-FHSS transceiver testbed for verification of relevant technologies; however, this study has two limitations. First, there is a lack of test results for integrated DtS-IoT system verification due to the absence of final application development for target use cases with relevant KPIs. Second, there is a significant gap between verification using actual operating LEO satellites and laboratory-level verification. For future studies, we consider the following two topics.

\textit{Development and integration of final application for environmental monitoring}: We are currently developing a marine climate prediction analysis system as a final application considering use cases for environmental monitoring. To this end, we intend to collect and store marine climate data such as water temperature and waves by mounting a developed satellite and ground integrated terminal on a marine buoy. The collected data is delivered to the gateway via LEO satellite payload, processed by NMS at the gateway, and accumulated on the server. Finally, the accumulated data is used to predict future ocean climate through machine learning in a big data analysis system.

\textit{In-orbit test of the DtS-IoT system}: Currently,the LEO satellite payload of the developed DtS-IoT system has been developed at the EM level, and system verification has been performed through laboratory tests on the ground. This system verification was performed through transmission tests on each user and feeder link. Therefore, additional DtS-IoT system functional and performance testing is needed to verify that it can satisfy the relevant KPIs of the target use case. Finally, we aim to build and launch an LEO satellite payload containing the LR-FHSS receiver with a flying model (FM) level and perform an in-orbit test of the DtS-IoT system with it.


\end{document}